\documentclass[a4paper, amsfonts, amssymb, amsmath, reprint, showkeys, nofootinbib, twoside, floatfix]{revtex4-1}
\usepackage[english]{babel}
\usepackage[utf8]{inputenc}
\usepackage[colorinlistoftodos, color=green!40, prependcaption]{todonotes}
\usepackage{caption}
\usepackage{subcaption}
\usepackage{amsthm}
\usepackage{mathtools}
\usepackage{physics}
\usepackage{xcolor}
\usepackage{graphicx}
\usepackage{adjustbox}
\usepackage{placeins}
\usepackage[T1]{fontenc}
\usepackage{lipsum}
\usepackage{csquotes}

\usepackage[pdftex, pdftitle={Article}, pdfauthor={Author}]{hyperref}
\usepackage{tikz,xcolor,hyperref}
\definecolor{lime}{HTML}{A6CE39}
\DeclareRobustCommand{\orcidicon}{
	\begin{tikzpicture}
	\draw[lime, fill=lime] (0,0) 
	circle [radius=0.16] 
	node[white] {{\fontfamily{qag}\selectfont \tiny ID}};
	\draw[white, fill=white] (-0.0625,0.095) 
	circle [radius=0.007];
	\end{tikzpicture}
	\hspace{-2mm}
}
\foreach \x in {A, ..., Z}{
	\expandafter\xdef\csname orcid\x\endcsname{\noexpand\href{https://orcid.org/\csname orcidauthor\x\endcsname}{\noexpand\orcidicon}}
}

\begin{document}
\title{Erasure of Strings and Vortexes}
\author{Gia Dvali}
\affiliation{
	Arnold Sommerfeld Center,
	Ludwig-Maximilians-Universit{\"a}t,
	Theresienstra{\ss}e 37,
	80333 M{\"u}nchen,
	Germany}
\affiliation{
	Max-Planck-Institut f{\"u}r Physik,
	F{\"o}hringer Ring 6, 80805 M{\"u}nchen,
	Germany}
\author{Juan Sebasti\'an Valbuena-Berm\'udez\orcidA{}}
\email{juanv@mpp.mpg.de}
 \affiliation{
 	Arnold Sommerfeld Center,
 	Ludwig-Maximilians-Universit{\"a}t,
 	Theresienstra{\ss}e 37,
 	80333 M{\"u}nchen,
 	Germany}
 \affiliation{
 	Max-Planck-Institut f{\"u}r Physik,
 	F{\"o}hringer Ring 6,
 	80805 M{\"u}nchen,
	Germany}
\date{\today} 
\begin{abstract}
The interaction of defects can lead to a phenomenon  of erasure.  During this process, a lower-dimensional object gets absorbed and dissolved by a higher-dimensional one. 
The phenomenon is very general and has a wide range of implications, both cosmological and fundamental. 
In particular, all types of strings, such as cosmic strings, QCD flux tubes, or fundamental strings, get erased when encountering a defect, either solitonic or a $D$-brane that deconfines their fluxes. 
This leads to a novel mechanism of cosmic string break-up, accompanied by gravitational and electromagnetic radiations. The arguments based on loss of coherence and the entropy count suggest that the erasure probability is very close to one, and strings never make it through the deconfining layer.
We confirm this by a numerical simulation of the system, which effectively captures the essence of the phenomenon: a $2+1$-dimensional problem of interaction between a Nielsen-Olesen vortex of a $U(1)$ Higgs model and a domain wall inside which the $U(1)$ gauge group is unHiggsed and the magnetic flux is deconfined.  In accordance with the entropy argument, in our simulation, the vortex never makes it across the wall. 
\end{abstract}

\maketitle
\section{Outline} \label{sec:outline}

Topological defects, such as domain walls, cosmic strings, and magnetic monopoles, appear in various quantum field theories. 
Their  classification is determined by the  topology of the vacuum manifold, $\mathcal{M}$. Due to their important role in cosmology, the topological defects have been objects of extensive studies; for a review, see~\cite{Vilenkin:2000jqa}. 

Another important aspect of topological defects is that they can serve as dual laboratories for studying the fundamental ``electric''-type extended objects, for which the description in terms of non-singular classical solutions is not known. 
The examples of such objects  are provided by QCD strings of a confining gauge theory, as well as by the fundamental strings ($F$-strings) and $D$-branes of string theory.    

In some cases, one and the same theory contains objects of different dimensionality that share the relevant vacuum manifold. For example, a confining gauge theory contains QCD strings. Strings can be arbitrarily long, provided no light quarks are included in the theory. However, the same theory can also contain domain walls that support a deconfined phase in their world-volume. This setup was proposed in Dvali-Shifman (DS) model~\cite{Dvali:1996xe} and has several implications. 
 
First, such domain walls localize a massless gauge field. Correspondingly, the QCD strings that carry the chromo-electric flux can end on them. The flux is spread inside the wall in the form of a Coulomb electric flux of the  $2+1$-dimensional world-volume theory. 

Notice that, due to gauge invariance, this general feature of the DS mechanism persists for the localization of massless gauge fields on the branes of higher dimensionality and therefore has important implications for ``brane-world'' scenarios~\cite{Dvali:1996xe, Arkani-Hamed:1998jmv}. 
In all these setups, the Standard Model gauge fields, such as  the $U(1)$-photon, are localized on a $3$-brane embedded in a higher dimensional space. This implies the existence of bulk strings in the form of flux tubes of the $U(1)$ electric field.

In this sense, the theory~\cite{Dvali:1996xe} realizes a setup very similar to $D$-branes of string theory.  Some aspects of this connection that will be extended in our analysis were studied in~\cite{Dvali:2002fi, Dvali:2007nm}.  
 
The connection with $D$-branes is also evident in the second realization of the deconfining domain wall discussed in~\cite{Dvali:1996xe}. This wall is formed by the gaugino condensate of a supersymmetric  $SU(N)$ gauge theory. It was shown there that, due to a central extension of the supersymmetric algebra,  the tension of the wall could be computed exactly in terms of the gaugino condensate, which in large-$N$ scales as $N$. The following observation by Witten also supports the connection to $D$-branes~\cite{Witten:1997ep}.  The $N$-scaling of the wall tension matches the expectations for a $D$-brane of the string theory with string coupling $1/N$ realized by large-$N$ QCD~\cite{tHooft:1973alw}. As studied in many subsequent papers, this matches the idea of a partial deconfinement of $SU(N)$ theory within the domain wall formed by the gaugino condensate.    
 
Correspondingly, the study of interactions between strings and deconfining domain walls of~\cite{Dvali:1996xe} and of their magnetic dual~\cite{Dvali:2002fi} can give glimpses of new dynamics of QCD strings in confining theories and also create some useful parallels with  their string-theoretic ``cousins''.   
 
Another framework of our interest is given by the domain walls and monopoles formed by the same order parameter.  
This was shown to be the case in the minimal grand unified theory~\cite{Dvali:1997sa}. 
In this case, the grand unified symmetry, which is spontaneously broken in the vacuum, is restored in the core of the domain wall.
In all the above cases, the interaction among the objects of different dimensionality can result in the ``erasure'' of the lower dimensional objects.
This term was coined in~\cite{Dvali:1997sa}, where it was shown that when a monopole encounters a wall, it unwinds, and the magnetic charge spreads over the wall. 

The original motivation of this proposal was to show that, even if the grand unified phase transition occurs after inflation, the cosmological monopole problem~\cite{Zeldovich:1978wj, Preskill:1979zi} can nevertheless be absent. 
This is due to the fact that besides monopoles, the same phase transition forms  unstable domain walls. The interaction between the monopoles and domain walls leads to monopole erasure and the subsequent decay of the domain walls. This mechanism has been further investigated and largely confirmed in the series of papers~\cite{Alexander:1999mf, Pogosian:1999zi, Brush:2015vda}. 
It was found that the monopole unwinds on the wall while the winding number spreads out on the surface.  

We also remark that an analogous erasure effect in the interaction of skyrmions with walls was detected in~\cite{Kudryavtsev:1997nw}. 

An important question concerns the probability of the erasure. 
An intuitive argument in~\cite{Dvali:1997sa} indicated that the probability of a monopole passage through the wall is highly suppressed.
The reason is the loss of coherence. 
When the monopole hits the wall, it sets traveling waves that take away part of the coherence required for recreating a monopole on the other side.

The coherence argument supporting the high probability of erasure was made more explicit in the recent work~\cite{Dvali:2022vwh}. There, the annihilation of a pair of `t Hooft-Polyakov monopoles connected by a string was studied.
This work extended the analysis beyond the approximation of zero widths of string and monopoles previously performed by Martin and Vilenkin~\cite{Martin:1996cp}. It was observed that in a head-on collision, the monopoles are never re-created. 
The loss of coherence and entropy suppression explained this.
Once the monopole-anti-monopole come on top of each other, the system loses coherence due to induced waves. From here, the system evolves towards the most entropic state. The highly coherent configuration of a monopole-anti-monopole pair connected by a string has very little entropy compared to the entropy of waves of gauge and Higgs particles.
 
Putting it differently, the state of monopole-anti-monopole has much lower entropy than what is required for saturating the unitarity bound~\cite{Dvali:2020wqi}. Correspondingly, the system chooses to decay into the waves rather than to recreate monopoles.  
It was further argued in~\cite{Dvali:2022vwh} that the same counting must hold in the case of annihilation of heavy quarks connected by a QCD string.

The entropy argument explaining the suppression of monopole recreation in~\cite{Dvali:2022vwh} must also apply to the case of the monopole collision with the wall. 
Mentally, we can think of a wall as a layer filled with a densely packed ``sea'' of monopoles and anti-monopoles with a zero net magnetic charge. When a monopole hits the wall, it annihilates with an anti-monopole. 
This process endows the wall with a magnetic charge which spreads over it. A further re-creation of the monopole on the other side of the wall is subjected to a similar entropy suppression as in the case of an annihilating monopole pair studied in~\cite{Dvali:2022vwh}.

In the present paper, we wish to enhance the domain of study of the phenomenon of erasure by including the interaction between a domain wall and a vortex/string.   
Apart from the immediate goal of extending the erasure mechanism to the vortex-wall system, several motivations exist. This model represents a dual prototype of the interaction between the QCD string and a domain wall of the deconfined phase of the DS model~\cite{Dvali:1996xe}.  

As said, the universal feature of the DS gauge field localization mechanism is: {\it confining bulk/deconfining brane}.   
Due to this, the same applies to generic brane-world models in which a massless photon is localized on a brane embedded in a higher dimensional bulk~\cite{Arkani-Hamed:1998jmv}. Therefore, all such models must possess cosmic strings in the form of bulk $U(1)$ flux tubes of confined photons.
Strings  are erased upon colliding with ``our'' brane.
This erasure provides a novel mechanism of string decay, with potentially observable signatures of gravitational and electromagnetic radiation. 
 
Likewise, a simple vortex/wall system also represents a prototype describing the erasure of a fundamental string by a $D$-brane~\cite{Dvali:2002fi}.  

Besides the benefit of understanding the $D$-brane/string dynamics from the effective field theoretic perspective, this study has cosmological implications for the evolution of networks of  $D$-strings and $F$-strings~\cite{Copeland:2003bj, Dvali:2003zj}.    
Such defects are produced after brane inflation~\cite{Dvali:1998pa, Dvali:2001fw} and thus can play an essential role in post-inflationary cosmology~\cite{Sarangi:2002yt, Jones:2002cv, Majumdar:2002hy}. 
Annihilating strings can be an important source of gravity waves and primordial black holes~\cite{Dvali:2021byy}. 
The same applies to erasing strings, which give different mechanisms of gravitational radiation.  
 
Evidence of the erasure phenomenon has also been recognized experimentally. The interactions of topological defects have been studied in $^3\text{He}$~\cite{Misirpashaev:91, Eltsov:1998fv}.
A-phase vortices and domain walls separating the A and B phases of $^3\text{He}$ have been investigated and observed experimentally.
It has been found that singular vortices do not penetrate from one phase into the other. The measurements show that the vortices experience a force from the advancing interface and are pushed as a vortex layer in front of it. 
The authors identified critical velocity, at which a vortex will leave the layer and penetrate through the interface, transforming thereby into a  new structure~\cite{KRUSIUS1994376}.
This behavior indicates that A-phase vorticity cannot cross the AB interface. Instead, it is accumulated on the A-phase side of the interface such that it coats the interface with a dense vortex layer~\cite{Finne:2006bi}.

In the present paper, we shall study the unwinding process of a Nielsen-Olesen Vortex/String~\cite{Nielsen:1973cs} during its collision with a Domain Wall containing a core with a Coulomb-like phase, inside which the $U(1)$ symmetry is unHiggsed.
We consider the previous dynamics in a minimal model that simultaneously allows the existence of vortices and domain walls.
Since, in the case of a planar string parallel to a domain wall, the problem is effectively $2+1$-dimensional, we shall study the dimensionally reduced model with domain walls and vortexes.
This reduction has a double advantage. On one hand, this model represents a simplified version of a wall-monopole system of $3+1$-dimensional theory. 
On the other hand, it effectively captures the $3+1$ dimensional dynamics of erasure of cosmic strings by walls (or $F$-strings by  $D$-branes) as long as the curvature radius of the string loop and the wall bubble is larger than the string/wall thickness. Figure \ref{fig:StringsOnWall} shows a graphical representation of such a situation. 

\section{Dvali-Shifman Model} 
First of all, we shall briefly review the model of~\cite{Dvali:1996xe}. 
We shall consider the version of the model with a single scalar field, as discussed in~\cite{Dvali:2002fi, Dvali:2007nm}.
Let us introduce a $SU(2)$ gauge theory with the Higgs field $\phi^{a}$  in the adjoint representation of the group, with $a=1,2,3$ the adjoint index. The Lagrangian has the following form,  
\begin{equation}
    \mathcal{L}=-\frac{1}{4}F_{\mu \nu}^aF^{\mu \nu a} +(D_\mu \phi)^a (D^\mu \phi)^a - V(\phi),
    \label{eq:LagrangianDS}
\end{equation}
where the potential is chosen as, 
\begin{equation}
    V(\phi)=\lambda^2(\phi^b\phi^b)(\phi^a\phi^a-\nu^2)^2.
    \label{eq:PotentialDS} 
\end{equation}
The gauge field strength is $F_{\mu\nu}^a=\partial_\mu A_\nu^a -\partial_\nu A_\mu^a - e\epsilon^{abc}A_{\mu}^bA^c_{\nu}$, and the covariant derivative is $(D_\mu \phi)^a=\partial_\mu \phi^a - e\epsilon^{abc}A_{\mu}^b\phi^c$. 

Notice that the non-renormalizable appearance of the potential is no concern for our analysis since such a potential can easily be obtained from the renormalizable theory of~\cite{Dvali:1996xe} by integrating out  an additional gauge singled field. Several such examples were discussed in~\cite{Dvali:2002fi}. 

At the classical level, this theory possesses two degenerate vacua. In the first minimum, the vacuum expectation value (VEV) of the Higgs triplet vanishes, $\phi^a=0$. Correspondingly, all three $SU(2)$-gauge bosons are massless. The Higgs triplet has a mass $m_{\phi} = \sqrt{2}\lambda\nu^2$.  

In the second vacuum, $\phi$ has a non-zero VEV, and $SU(2)$ is  Higgsed down to $U(1)$. 
In the basis in which the VEV is chosen as $\phi^a = \delta^{a3}\nu$, the massless $U(1)$ gauge field is $A_{\mu}^3$,  whereas the other two components form a charged massive vector field and its anti-particle, $A^{\pm}_{\mu} \equiv \frac{1}{\sqrt{2}}(A^1 \pm iA^2)$.
The longitudinal components of these massive vectors are the two would-be Goldstone components of the Higgs triplet, $\phi^{\pm}_{\mu} \equiv \frac{1}{\sqrt{2}}(\phi^1 \pm i \phi^2)$. The third component is a neutral scalar of mass $m_h=2 \sqrt{2} \lambda v^2$.
 
Notice that the Higgs vacuum, $SU(2) \rightarrow U(1)$, supports the 't Hooft-Polyakov magnetic monopoles in the form of solitons.
These are in the ``magnetic'' Coulomb phase.   
At the same time, the electric charges are in the ``electric'' Coulomb phase.   

At the quantum level, the situation changes in the following way. 
In the unHiggsed vacuum, the theory confines and develops a mass gap set by the corresponding scale of QCD, $\Lambda$.
At distances larger than $\Lambda^{-1}$, the theory becomes a theory of composite degrees of freedom. 
The chromo-electric field is trapped in flux tubes which form the QCD strings. 
This phenomenon is commonly understood as the dual Meissner effect of condensation of $SU(2)$ magnetic monopoles~\cite{tHooft:1981bkw}.

In the Higgs vacuum, the situation is different. If $\nu \gg \Lambda$, the Higgs phase is unaffected by the quantum effects.  Then, at distances $> 1/\nu$, the low energy theory is described by a massless $U(1)$-photon and thereby is in the Coulomb phase. 
 
The situation can thus be summarized as follows,  
\begin{equation}
    \left<\phi^a\right>=\left\{ \begin{array}{llc}
             0,   &\textbf{SU(2) confining phase,}\\
             \\ \delta^{a3} \nu, &\textbf{U(1) Coulomb phase.}
             \end{array}\right.
             \label{eq:VEVDS}
\end{equation}
In the $U(1)$-vacuum,  both elementary electric charges and solitonic magnetic ones are in the Coulomb phase. 
In the $SU(2)$ confining vacuum, the magnetic charges condense, and the magnetic field is screened, whereas the electric field is trapped in fluxed tubes, and electric charges are confining.

The two vacua can coexist while being separated by a domain wall. 
The classical solutions for  infinite wall and anti-wall located in $x=0$ plane  can be evaluated exactly and have the following forms \footnote{
   See also~\cite{Gani:2014gxa} }
   \begin{align}
\phi_{(\pm\nu,0)}(x)&=\pm\nu  \sqrt{\frac{1}{1+e^{m_h
x}}},\label{eq:(nu,0)DS}\\
\phi_{(0,\pm\nu)}(x)&=\pm\nu  \sqrt{\frac{1}{1+e^{-m_h
x}}},\label{eq:(0,nu)DS}
 \end{align}

In this situation, one can envisage the following dynamics. The massless photon existing in $U(1)$-vacuum cannot pass through the wall unless its energy exceeds the mass gap of the confining phase.  
Hitting the wall with a high-energy photon, we can excite some glueballs on the other side.  

\begin{figure*}
\centering
    \includegraphics[width=0.8\textwidth]{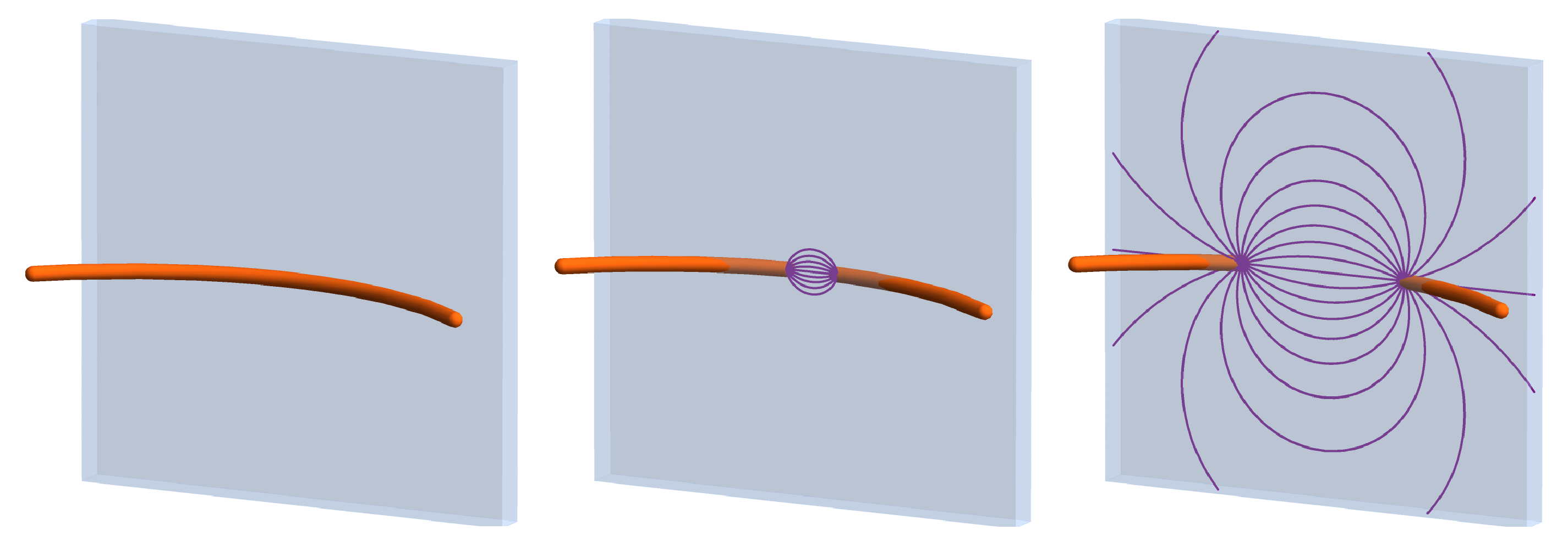}
    \caption{Graphical representation of the interaction between a string of the bulk theory, in orange, with a layer representing a domain wall or a brane, in gray. Whenever the string touches the layer,  the flux flowing through it spreads throughout the layer in the form of a $U(1)$ Coulomb electric field, in purple. From the point of view of a $2+1$-dimensional observer, the string-wall junction plays the role of a point-like electric charge. The same behavior applies to the dual case.}
	\label{fig:StringsOnWall}
\end{figure*}

At the same time, a glueball of energy $<\nu$ hitting the wall from the  $SU(2)$- confining side will excite massless photons propagating on the other side of the wall. 
Of course, in both cases, there also will be waves induced along the wall.  
In this setup, in~\cite{Dvali:2007nm}, the DS model was realized by  placing the parallel wall and anti-wall at some distance $l$ apart. 
In this way,  a layer of the $U(1)$-Coulomb phase of thickness $l$ is created in between the two confining $SU(2)$-phases. 
In this case, the effective $2+1$-dimensional theory describing the physics inside the layer at distances $> l$ contains a massless photon~\cite{Dvali:1996xe}. 
This theory confines, due to instantons representing the monopoles that tunnel through the layer~\cite{Dvali:2007nm}.
In this sense, the low energy $2+1$ dimensional theory realizes the confinement mechanism by Polyakov~\cite{Polyakov:1976fu}.
However, the scale of the confinement of the world-volume $U(1)$-theory is exponentially small and can be ignored for present purposes. 
Correspondingly, the system supports the QCD strings of the bulk $SU(2)$ theory that end on the layer. 
Whenever the  QCD string touches the layer, the flux flowing through the string spreads throughout the layer in the form of a $U(1)$ Coulomb electric field. From the point of view of a  $2+1$-dimensional observer, the string-wall junction plays the role of a point-like electric charge, as shown in Figure \ref{fig:StringsOnWall}.

In this way, the model realizes the idea of a field-theoretic $D$-brane.
The QCD strings attached to the layer play the role of the fundamental open strings, whereas the layer takes up the role of a $D$-brane.
Due to the above, a QCD string can be erased if it encounters a layer. 
What is the probability of erasure? That is, we would like to know what are the chances for a QCD string to pass through the layer.

For definiteness, we consider a string that lies in the plane parallel to the layer.
For example, the simplest case is a straight infinite string parallel to the layer.     
Of course, if the string moves slowly, the erasure is inevitable. 
The question is what happens if the sting moves sufficiently rapidly.
The coherence-loss argument suggests that such a passage must be exponentially suppressed. 
Namely, upon the collision with the layer, the QCD string is expected to create the traveling waves which take away coherence. This suppresses the possibility of creating a highly coherent configuration on the other side of the layer. String does not have sufficient entropy for matching the entropy of the multi-particle waves.
 
We shall study this phenomenon on a dual system in which the role of the QCD string is played by a $U(1)$-magnetic flux-tube of Nielsen-Olesen type. 
In the case of a string parallel to the wall, due to translation symmetry along it,  the problem effectively becomes $2+1$-dimensional. The same is true in any collision process in which the curvature radii of touching wall and string segments are much larger than the thicknesses of the two objects.  
    
In all such situations,  the problem can be effectively substituted by a dimensionally reduced version in which a vortex replaces the string. As already noticed, since the vortex is point-like, the $2+1$-dimensional version also serves as a prototype model for the erasure of a monopole by the domain wall. 
  
\section{The Model}
We shall study the dimensionally reduced version of the $U(1)$ model constructed in~\cite{Dvali:2002fi} as a dual prototype of $SU(2)$ confining theory (\ref{eq:LagrangianDS}). 
It is obtained by replacing the  $SU(2)$ gauge group by $U(1)$, and replacing the adjoint scalar Higgs of $SU(2)$ by a complex  scalar charged under $U(1)$.
Thus, let us consider a $(2 + 1)$-dimensional renormalizable model of a complex scalar field, $\phi$ with a $U(1)$ gauge symmetry, and Lagrangian given by,
\begin{equation}
    \mathcal{L}=-\frac{1}{4}F_{\mu \nu}F^{\mu \nu}+(D_\mu \phi)^* D^\mu \phi - V(\phi),
    \label{eq:Lagrangian}
\end{equation}
where the potential is
\begin{equation}
    V(\phi)=\lambda^2\phi\phi^*(\phi\phi^*-\nu^2)^2.
    \label{eq:Potential} 
\end{equation}
The gauge field strength is $F_{\mu\nu}=\partial_\mu A_\nu -\partial_\nu A_\mu$, and the covariant derivative is $D_\mu=\partial_\mu - ieA_\mu$.  Notice that the form of the potential is similar to (\ref{eq:PotentialDS}). 
This allows having 
the two degenerate vacua with different phases of the gauge theory, with and without a mass gap. 
The gapless phase is again represented by a massless $U(1)$-gauge field in the Coulomb regime. 
However, unlike the previous example, the phase with a mass gap is realized differently: instead of becoming a subgroup of a confining $SU(2)$-theory, the  $U(1)$ is simply Higgsed. 

The field equations for $\phi$ and $A_\mu$ are respectively
\begin{equation}
    \Box \phi + \frac{\partial V(\phi)}{\partial \phi^*}=0,
    \label{eq:eom-phi}
\end{equation}
\begin{equation}
    \partial_\mu F^{\mu \nu}=j^\nu,
    \label{eq:eom-A}
\end{equation}
where $\Box=D_\mu D^\mu$, and is $j^\mu=-ie\left(\phi^*D^\mu \phi-(D^\mu \phi)^*\phi\right)$ is the Noether current associated to the $U(1)$ symmetry.

As in the $SU(2)$ example, the minimum of the potential (\ref{eq:Potential}) has two disconnected components. As the field $\phi$ acquires a certain VEV, $\left<\phi\right>$, it has the following two possibilities:
\begin{equation}
    \left<\phi\right>=\left\{ \begin{array}{llc}
             0,   &\textbf{Coulomb or Symmetric Phase,}\\
             \\ \nu e^{i \alpha}, &\textbf{Higgs or Broken Phase,}
             \end{array}\right.
             \label{eq:VEV}
\end{equation}
where $\alpha=\alpha(x)$ is an arbitrary phase.

In the first possibility, $\left<\phi\right>=0$,  if we consider perturbations around the VEV, we find the following spectrum of excitations: a charged boson--corresponding to the field $\phi$--with mass $m_{\phi} = \lambda\nu^2$, and a massless gauge boson--corresponding to the gauge field $A_\mu$.
We notice that the spectrum corresponds to a Coulomb or symmetric phase.  
On the other hand, $\left<\phi\right>=\nu e^{i\alpha}$ corresponds to a Higgs or broken phase. More precisely, the Higgs mechanism occurs, and the gauge boson becomes massive. 

Lets consider perturbations about the VEV as $\phi=\left(\nu + \frac{h}{\sqrt{2}}\right)e^{i\theta}$, where $h$ and $\theta$ are real fields.
We then find the  following spectrum: a neutral Higgs boson--corresponding to the scalar field $h$--with mass $m_h=2\lambda\nu^2$; and a massive vector boson--corresponding to the vector field $B_\mu=A_\mu-\frac{1}{e}\partial_\mu \theta,$ with mass $m_v= \sqrt{2}e\nu$.

So far, we have shown that there exist two different phases in the  model. 
For the case at hand, there can exist static field configurations that depend on one space dimension and asymptotically approach the two different phases. 
We will refer to these configurations as $(\nu,0)$-domain walls. 
Additionally, in the broken phase, one can show that  $\pi_1(\mathcal{M}_{\text{H}})=\mathbb{Z}$. Thus, there exist static field configurations that approach the Higgs Phase with non-zero \textit{winding number} asymptotically. These configurations are Nielsen-Olesen-like Vortices.  

It is possible to form a finite-size configuration that asymptotically interpolates two broken phases with a core inside which the full symmetry group $U(1)$ is nearly unHiggsed.
Strictly speaking, such configurations are not topologically protected. 
However, they can be sufficiently long-lived for our purposes. 
To set the terminology, we will refer to this configuration as a \textit{Coulomb vacuum layer}. This name encodes the information that in the limit of an infinite layer width, the electric and magnetic charges would be in the Coulomb phase. 
However, for a finite width, this is not the case, as shall be explained in a separate section. So, the name ``Coulomb'' refers to the nature of the layer in the limit of infinite width. 
 
As a first approximation, a Coulomb vacuum layer can be achieved as a concatenation of two $(\nu,0)$-domain walls: one interpolating between the Higgs and the Coulomb phases and a second one interpolating between the Coulomb and the Higgs phases.
As a result, inside the core of a Coulomb vacuum layer, the $U(1)$ symmetry is nearly unHiggsed, as required, while asymptotically interpolating towards Higgs phases. 

We are then ready to study the interaction of vortices and Coulomb vacuum layers.
In the following sections, we describe in detail the Coulomb vacuum layers, the vortices, and their interactions. 
Finally, we present our results of the numerical simulations and discuss how the vortex unwinding occurs in the core of the Coulomb Vacuum layer. 

\subsection{Domain Walls and Vacuum Layers}
Domain walls and vortices belong to the spectrum of the model (\ref{eq:Lagrangian}). 
A $(\nu,0)$-domain wall is a topological field configuration interpolating the Higgs and the Coulomb  vacua.
More precisely, let's assume for now that the field $\phi=\phi(t,x,y)$ depends only on the coordinate $x$. 
Moreover, let's assume that the domain wall does not carry an electric charge. 
Therefore, we can fix $A_\mu=0$ and $\phi$ to be real.
Under these assumptions, the Lagrangian (\ref{eq:Lagrangian}) becomes the (1+1)-dimensional $\phi^6$  model.
The Field profiles can be computed analytically to be
 \begin{align}
\phi_{(\pm\nu,0)}(x)&=\pm\nu  \sqrt{\frac{1}{1+e^{m_h
x}}},\label{eq:(nu,0)dw}\\
\phi_{(0,\pm\nu)}(x)&=\pm\nu  \sqrt{\frac{1}{1+e^{-m_h
x}}},\label{eq:(0,nu)dw}
 \end{align}
and we refer to them as  $(\pm\nu,0)$ and $(0,\pm\nu)$-domain walls, respectively.
There are four different solutions corresponding to the possible asymptotic behaviors of a Domain Wall: $(\nu,0)$, $(-\nu,0)$, $(0,\nu)$, and $(0,-\nu)$. \footnote{Here we used the notation $(\phi_{-},\phi_{+})$, where $\phi_{\pm}\equiv \lim_{x\to\pm\infty}\phi(x)$.}. 

\begin{figure}
    \centering
    \includegraphics[width=0.4\textwidth]{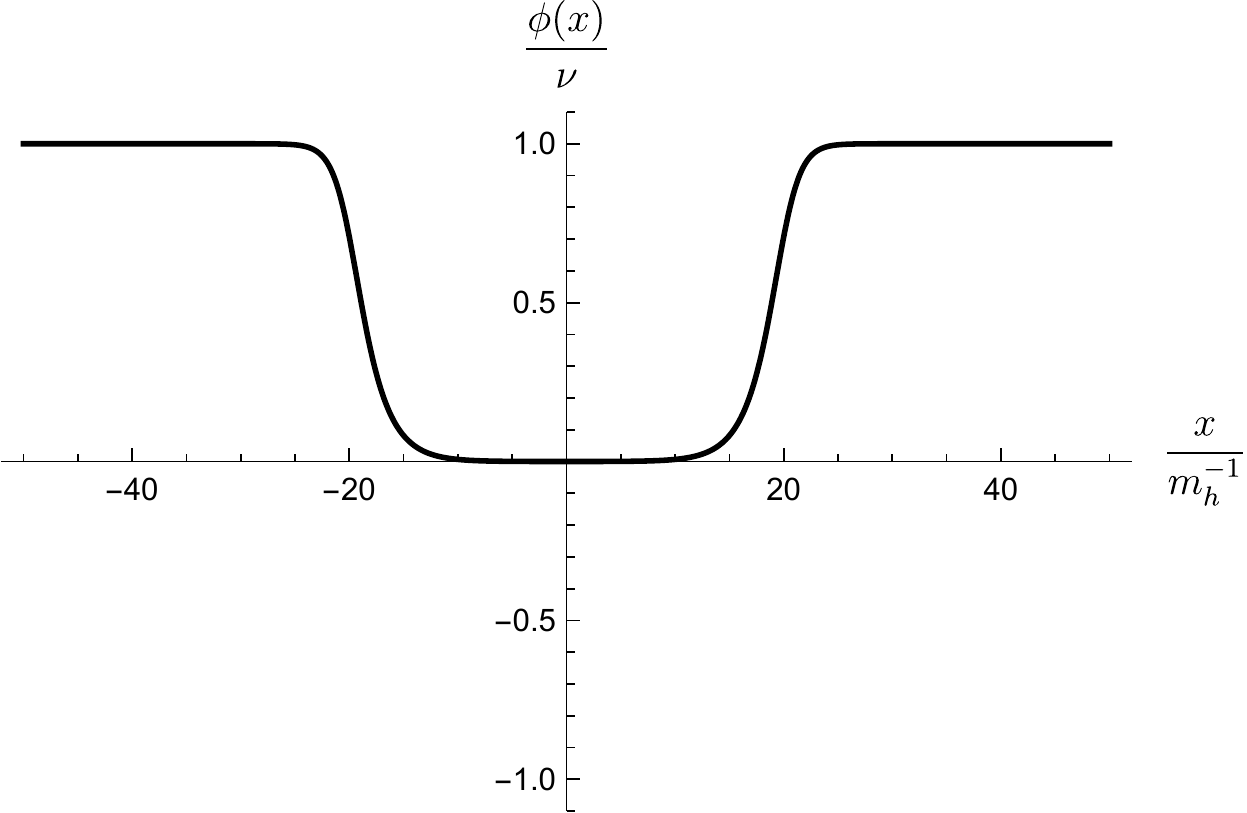}
    \caption{Coulomb Vacuum Layer profile,  $\phi_{(\nu,0,\nu)}(x)$, for a core size 
    $l=40m_h^{-1}$}
    \label{fig:(nu,0,nu) DW-Profile}
\end{figure}

Now, let us construct a field configuration interpolating between two different Higgs phases with a core inside which the symmetry is restored, i.e. a Coulomb Vacuum Layer.
We achieve this by concatenating a $(\pm\nu,0)$ and a $(0,\pm\nu)$-Domain Wall. The ansatz for these configurations is given by:
\begin{equation}
    \phi_{(\nu,0,\nu)}(x)=\phi_{\left(\nu,0\right)}\left(x+\frac{l}{2}\right)+\phi_{(0,\nu)}\left(x-\frac{l}{2}\right),
    \label{eq:DW-Ansatz}
\end{equation}
where $l$ is the distance between the domain walls. We will refer to these field configurations as a Coulomb Vacuum Layer or a $(\pm\nu,0,\pm\nu)$-domain wall.
Figure \ref{fig:(nu,0,nu) DW-Profile} shows an example of a Coulomb Vacuum Layer profile.

Notice that the ansatz (\ref{eq:DW-Ansatz}) is not a solution to the static field equation (\ref{eq:eom-phi}). Therefore, it is expected to be unstable under time evolution. 
The dynamical evolution of such configurations has been investigated by Gani et al. in~\cite{Gani:2014gxa}. 
They used the collective coordinate approximation and numerical simulations to study the classical evolution of the ansatz. 
There are two conclusions that are relevant to our discussion.
The first one is with respect to the dynamical evolution of the $(\nu,0,\nu)$-domain wall.
It is energetically favorable for the $(\nu,0)$ and the $(0,\nu)$-Walls to attract each other, and eventually, they collide. 
As a consequence, the Coulomb Vacuum Layer is unstable. 
The two colliding Walls can form a long-lived bound state (referred to as bion).
We have reproduced these phenomena by solving the field equation (\ref{eq:eom-phi}) numerically. The second relevant conclusion is with respect to the dynamical evolution of the $(-\nu,0,\nu)$-domain wall. 
In this case, the two Walls repel each other till they get a limited speed, and the parameter $l$ increases infinitely\footnote{A similar discussion applies to the evolution of the $(\nu,0,-\nu)$- domain wall. We reproduced these phenomena numerically in both cases.}.

In this setup, a $(\nu,0,\nu)$-domain wall configuration forms a Coulomb vacuum layer characterized by its width $l$.
Although this configuration is unstable, we find numerically that if $40 m_h^{-1}\lesssim L$, the layer can be considered to be long-lived for time scales of order $\mathcal{O}(10^2m_h^{-1})$, where $m_h$ is the mass of the Higgs-like boson.

\subsection{Vortex}
It is well known that vortex lines--or cosmic strings--arise in models in which the vacuum manifold $\mathcal{M}$ is not simply connected. This is the case for the model (\ref{eq:Lagrangian}) in the Higgs phase.
In this case  $\pi_1(\mathcal{M}_\text{H})=\pi_1(U(1))=\mathbb{Z}$.  Each homotopy class corresponds to a different winding number, $n$. These solutions are similar to the Nielsen-Olesen~\cite{Nielsen:1973cs} vortex lines that arise in the abelian-Higgs model.
Moreover, the winding number is a topological number characterizing the field configuration, and it is an integral of motion~\cite{book:Rubakov:2002fi}.
In fact, the winding number can be written as
\begin{equation}
    n=\lim_{r\to\infty}\frac{1}{2\pi i \nu^2} \oint_{C_r} d x^i \frac{1}{2}\left( \phi^*\partial_i\phi-\phi\partial_i\phi^*\right),
    \label{eq:winding_Phi}
\end{equation}
where $C_r$ is the circle of radius $r$, and centered at the origin. The gauge invariance is explicit. 
Consequently, field configurations with a fixed winding number $n$  are (asymptotically) equivalent up to smooth gauge transformations. 
The winding number can also be computed from the flux of the magnetic field $B$ as
\begin{equation}
    n=\frac{e}{2\pi}\int B d^2x.
    \label{eq:winding_B}
\end{equation}

We  look for field configurations with finite energy, that asymptotically approach the Higgs Phase, and wind around $\mathcal{M_H}$ $n$ times.
As a first approach, we look for static cylindrical-symmetric solutions $(\phi, A_\mu)$ in the temporal gauge $A_0=0$. The most general--up to gauge transformations--invariant ansatz for the field profiles are:
\begin{align}
\phi(r,\theta)&=\nu e^{i n \theta} F(r),\label{eq:ansatz-phi}\\
A_i(r,\theta)&=-\frac{n}{er}\epsilon_{ij}n_j A(r),\label{eq:ansatz-A}
\end{align}
where $F(r)$, and $A(r)$ are smooth numerical functions that have the following asymptotic behavior:
\begin{equation}
F(r)\to 1, \, A(r)\to 1.
\end{equation}
Moreover, the requirement that the fields are smooth at $0$ implies that:
\begin{equation}
    F(0)=0, \,  A(0)=0.
\end{equation}
\begin{figure}
	\centering
	\includegraphics[width=0.4\textwidth]{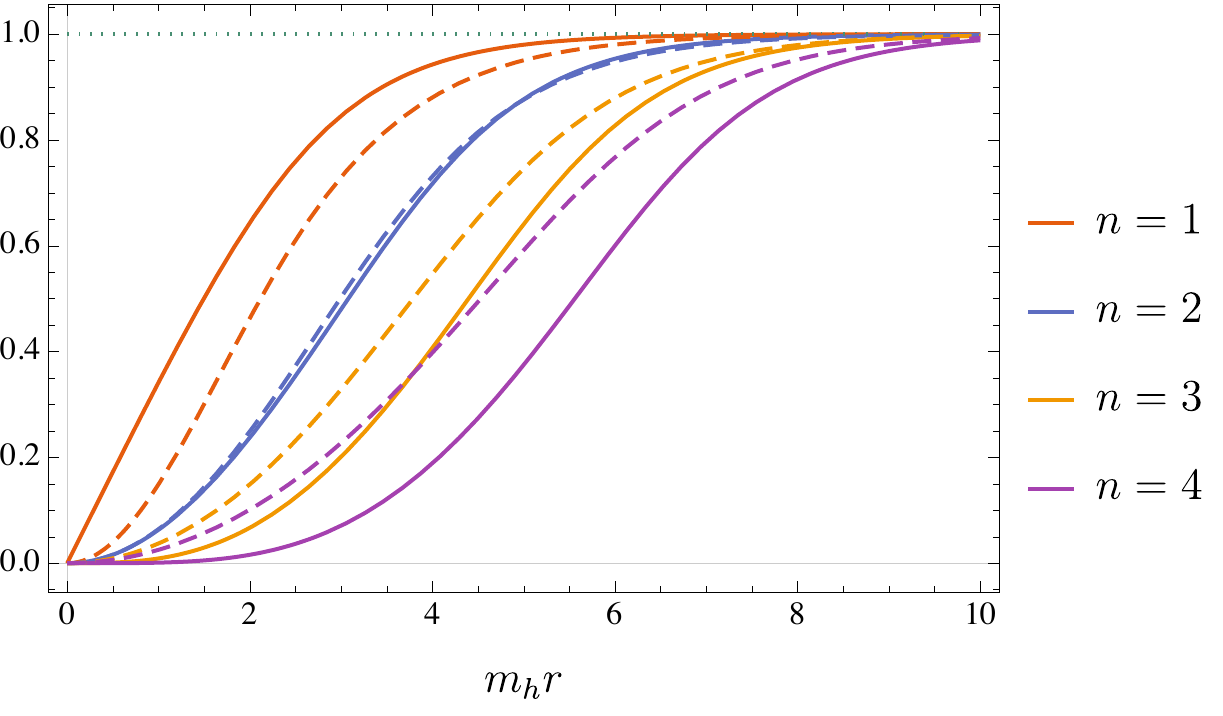}
	\caption{Numerical approximations of the Vortex profile functions $F(r)$, continuous lines, and $A(r)$, dashed lines, for different winding number, $n$. Here  $m_h=1$ and $m_v=1$.} 
	\label{fig:Vortex_Profile_m_h=m_v}
\end{figure}
Substituting the ansatz (\ref{eq:ansatz-phi}), and (\ref{eq:ansatz-A}) in the field equations (\ref{eq:eom-phi}), and (\ref{eq:eom-A}), we find the following system of equations for $A(r)$ and $F(r)$: 
\begin{equation}
\begin{aligned}
0=  &-r F''(r)-F'(r)+\frac{(1-A(r))^2}{r} n^2 F(r)\\
    &\ +\frac{m_h^2 }{4} r F(r) \left(F(r)^2-1\right)\left(3 F(r)^2-1\right),
\end{aligned}
\label{eq:Vortex-Profile-F}
\end{equation} 
\begin{equation}
0=-\frac{A''(r)}{r}+\frac{A'(r)}{r^2}-\frac{m_v^2 }{r}(1-A(r)) F(r)^2.
\label{eq:Vortex-Profile-A}
\end{equation}

Solutions to the equations (\ref{eq:Vortex-Profile-F}) and (\ref{eq:Vortex-Profile-A}) determine the field profiles (\ref{eq:ansatz-phi}) and (\ref{eq:ansatz-A}).
Analytical solutions for $F$ and $A$ are not known so far, but approximate solutions can be found numerically.
In what follows we have set $\nu = 1$ unless stated otherwise.
We used a shooting parameter method to solve the equations (\ref{eq:Vortex-Profile-F}) and (\ref{eq:Vortex-Profile-A}). Examples of the solutions we found are shown in Figure \ref{fig:Vortex_Profile_m_h=m_v}.

\begin{figure}
\centering
\includegraphics[width=0.4\textwidth]{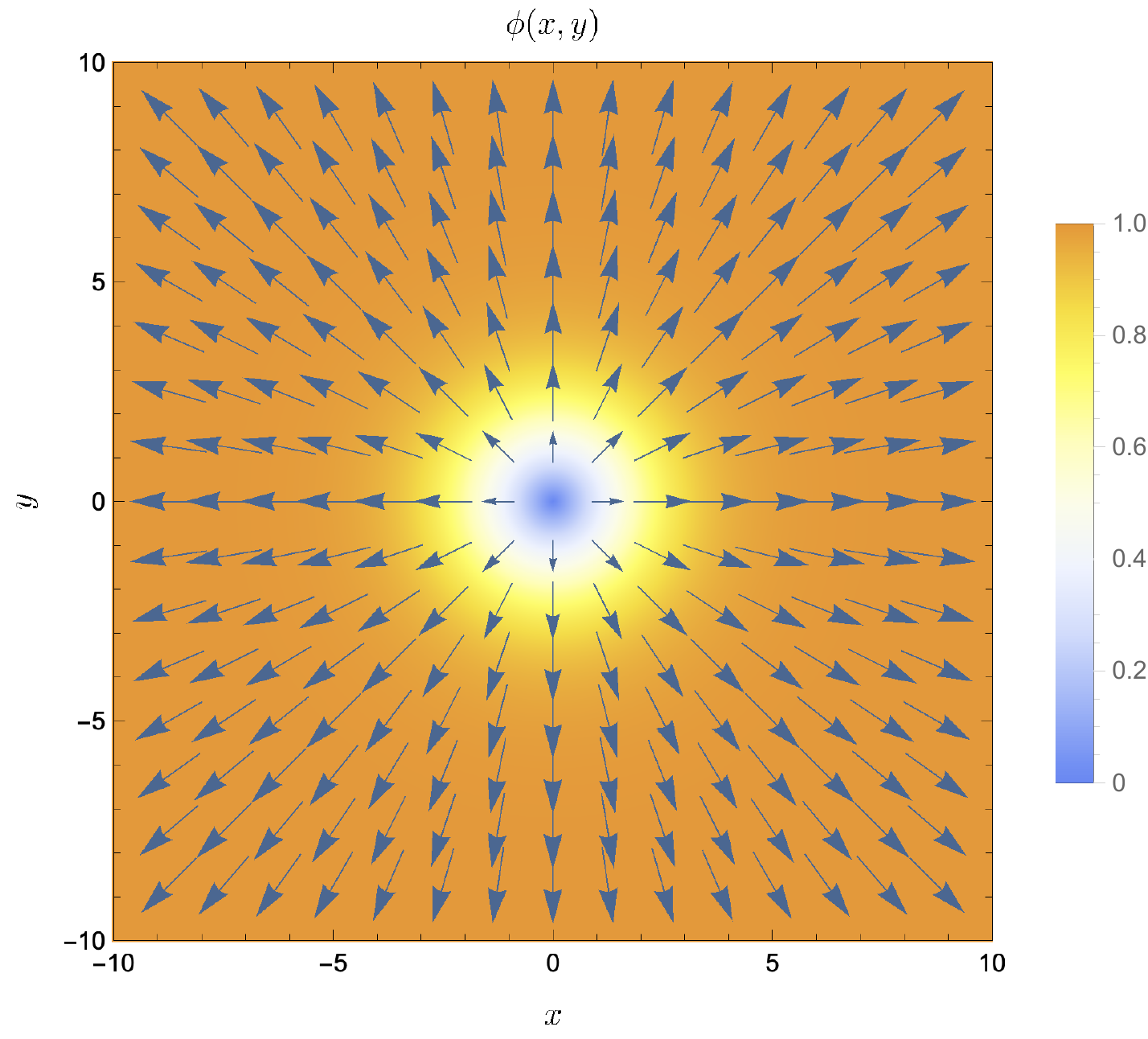}
 \caption{Vortex field configuration for the field $\phi$--equation (\ref{eq:ansatz-phi}). Here $n=1$, $m_h=1$, $m_v=1$, and $n=1$.   The vector at the point $(x,y)$ has components $\left(\text{Re}(\phi(x,y)),\text{Im}(\phi(x,y))\right)$. Consequently, the direction of each vector represents the phase $Arg(\phi(x,y))$, while the color represents the norm $|\phi(x,y)|$.} 
 \label{fig:CauchyData-Vortex-Phi}
\end{figure}

\begin{figure}
\centering			
\includegraphics[width=0.4\textwidth]{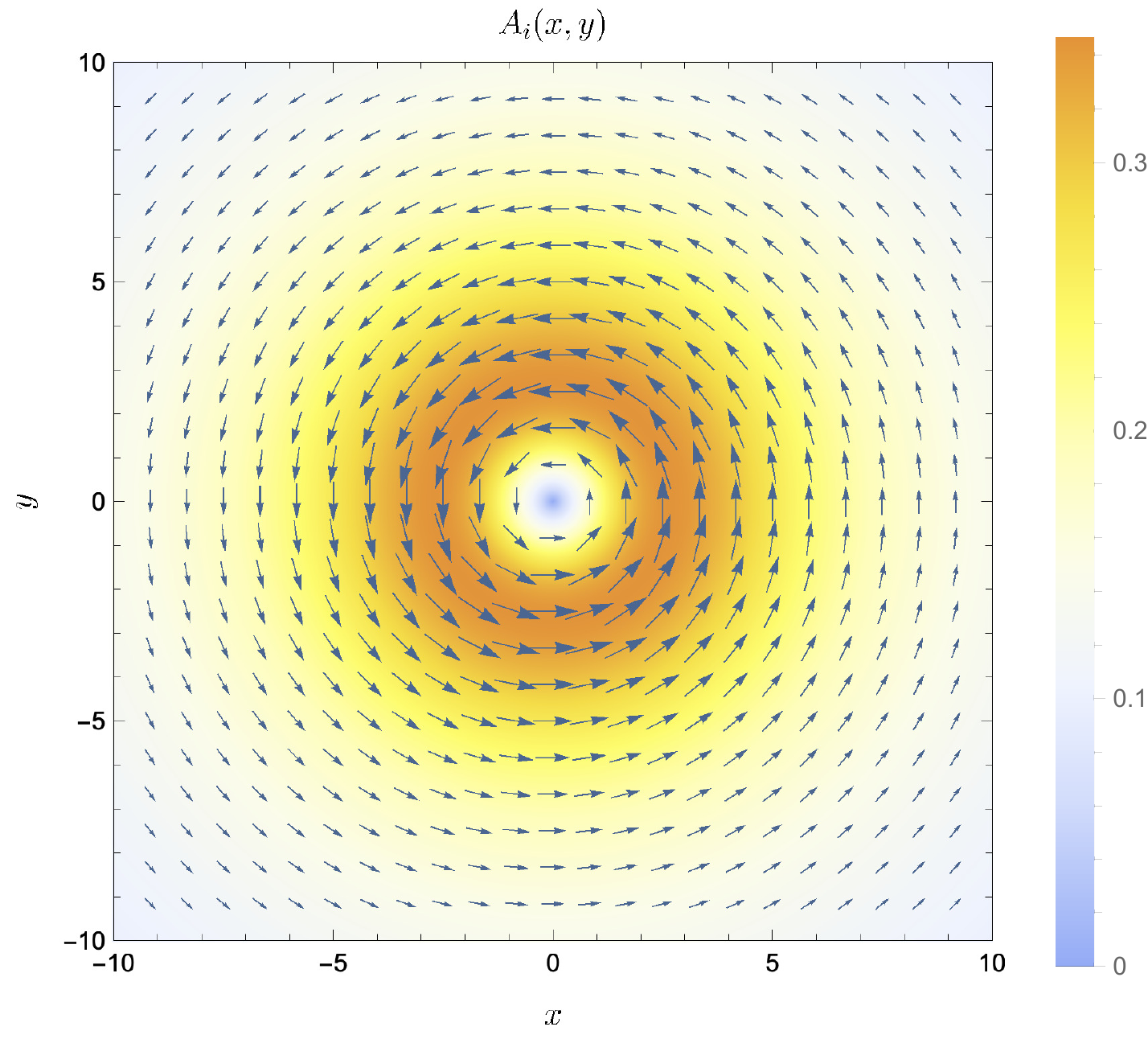}
\caption{Vortex field configuration for the field $A_i$--equation (\ref{eq:ansatz-A}). Here $e=1/\sqrt{2}$, $m_h=1$, $m_v=1$, and $n=1$. The vector at the point $(x,y)$ has components $(A_1(x,y),A_2(x,y))$. The color represents the norm $|A_{i}(x,y)|$} 
 \label{fig:CauchyData-Vortex-A}
\end{figure}

Figures \ref{fig:CauchyData-Vortex-Phi} and \ref{fig:CauchyData-Vortex-A} show the vortex profiles for $\phi$ and $A_\mu$, respectively. Using these field configurations as the initial conditions of the fields, and $\partial_t A_i=0$ and $\partial_t \phi=0$ as initial conditions of the time derivative of the fields, we were able to establish the numerical stability of such solutions.

\subsection{Phase Inside the Coulomb Layer} 
Let us briefly clarify certain aspects of the physics inside the Coulomb layer.  
As already said, we use this name to indicate that in the limit of infinite width, $l \rightarrow \infty$, the layer becomes a vacuum in which the gauge $U(1)$ theory is in the $2+1$-dimensional Coulomb phase. 
For example, the  electric potential energy of a pair of opposite charges separated by a distance $r$ scales as $\ln(r)$.   
In this vacuum, the magnetic vortex configurations are unstable.  
If we prepare a localized magnetic field with nonzero flux, it will spread out.   
Of course,  the same is true in  the $3+1$-dimensional version of the theory: the magnetic flux tubes are unstable in the Coulomb vacuum. 

In the Higgs vacuum, the picture is very different. First,  the electric 
 field potential decays exponentially at distances  $r> m_V$. 
In other words, in the Higgs phase, the electric charges become screened.   Secondly,  in the Higgs vacuum, the stable vortex configurations of the Nielsen-Olesen type exist with the quantized magnetic flux (\ref{eq:winding_B}). 
 Similarly, the theory uplifted to $3+1$-dimensions,
supports the vortex lines/cosmic strings.  

Let us now discuss the layer of a would-be Coulomb vacuum of a finite extend, $l$, separating the two Higgs vacua. 
This is described by a configuration of the type given in Figure \ref{fig:(nu,0,nu) DW-Profile}.   
Although naively, one may think that, if placed inside the layer,  the electric charges can be in the $1+1$-dimensional Coulomb phase, this is not the case.
 
It is easy to see that in the effective $1+1$-dimensional theory of the  layer,  no massless photon excitation exists but only the massive ones.
That is, a linearized gauge theory on the background shown in Figure \ref{fig:(nu,0,nu) DW-Profile}, has no gapless correlators\footnote{Of course, in $1+1$ dimensions there exists no propagating massless photon in general but there can exist a long-range electric field.  It is the absence of this field that is the key in the present discussion.}.

This is a manifestation of a well-understood effect
~\cite{Dvali:1996xe},~\cite{Arkani-Hamed:1998jmv}  that a massless ``electric'' photon cannot be localized  on a lower dimension defect (brane) by means of a Higgs mechanism in the bulk. 
The easy way to understand this is to remember that in the Higgs phase the vacuum is a superconductor with respect to $U(1)$ and the mobile charges are available. 
In the present case, this mobility is provided by the would-be Goldstone phase $\theta$. Due to this,  whenever a non-zero charge is placed inside the layer, the mobile bulk charges swiftly respond and create an infinite number of images that screen the electric field. 
In other words, the Higgs medium does not repel the electric field lines but  instead absorbs them.  
This is the reason why DS mechanism~\cite{Dvali:1996xe} of localization of the massless photon on the brane requires that the extra-dimensional bulk is in a confining phase rather than in a Higgs one. 
  
As already mentioned, in DS theory, at the quantum level, the $U(1)$ electric field of the low-dimensional theory is confining due to the monopole tunneling through the layer~\cite{Dvali:2007nm}.  That is, the low-dimensional theory of DS brane realizes the confinement via the mechanism of Polyakov~\cite{Polyakov:1976fu}.
  
In the dual version of the theory, a similar effect takes place with respect to the magnetic field.  
This is true both in $3+1$-dimensional theory as well in the dimensionally reduced $2+1$-dimensional one, which we are currently considering.  
 
That is, in both cases, if the vortex is placed inside the layer, the magnetic field will not spread all the way to infinity but will become confined at a finite distance.
In~\cite{Tetradis:1999tn}, this was explained by Tetradis as a close analogy with the Josephson Junction effect familiar in ordinary superconducting materials. 
 
Correspondingly, in our $1+1$ dimensional layer, the magnetic field is expected  to spread to a finite distance.
However, for $l > 1/m_V$,  this distance is exponentially large and therefore does not affect the probability of the erasure significantly. 
A similar effect is expected in the theory lifted to one dimension higher.
There too, an infinite straight cosmic string,  parallel to the wall, when erased by it, will spread the flux over an exponentially large distance.   

\section{Vortex Unwinding}
We are now ready  to discuss the erasure of a vortex by a Coulomb Vacuum layer sweeping. We simulated the collision for different parameters of the model. Within this approach, one can observe how the collision leads to the  unwinding of the scalar field, and the dissipation of the magnetic flux in the core of the Coulomb Vacuum Layer.  As a result, the magnetic field spreads and dissipated in the core of the layer.

To avoid Gauge redundancies, we use the time gauge, $A_t = 0$, as it is convenient for solving time evolution problems and numerically solving the field equations (\ref{eq:eom-phi}) and (\ref{eq:eom-A}). For the time gauge, the  \textit{Cauchy data}--$(A_i,\ \partial_t A_i,\ \phi,\ \partial_t \phi)$--or initial conditions must satisfy the Gauss constraint
\begin{equation}
    D_{i} F^{i0}=j^0
\label{eq:Gauss_Constraint}
\end{equation}
at $t=0$. Then, the field equations can be integrated to evaluate $A_i$ and $\phi$ at $t>0$. Proceeding in this way, after fixing the time gauge, the field equations (\ref{eq:eom-phi}) and (\ref{eq:eom-A}) become:
\begin{equation}
\begin{split}
\partial_t^2\phi = \
    &\partial_i\partial_i\phi- 2 i e A_i  \partial_i\phi  
    -\left[e^2 A_i A_i +i e \partial_i A_i \right]\phi\\
    &\hphantom{\partial_i\partial_i\phi }
    -\left[\lambda ^2 \nu ^4-4 \lambda ^2 \nu ^2\left| \phi  \right| ^2+3 \lambda ^2 \left| \phi \right| ^4\right] \phi,\\
\partial_t^2 A_x = \
    &\partial_y^2A_x -\partial_x\partial_y{A_y}
    -2 e \left| \phi  \right|^2 A_x 
    +2 \text{Im}\left[\phi  ^*\partial_x\phi   \right],\\
\partial_t^2 A_y = \
    &\partial_x^2A_y -\partial_y\partial_x{A_x} 
    -2 e \left| \phi  \right|^2 A_y 
    +2 \text{Im}\left[\phi  ^*\partial_y\phi   \right].\\
\end{split}
\label{eq:eom-Tgauge}
\end{equation}

The integration of the system of partial differential equations (\ref{eq:eom-Tgauge}) gives the time evolution for a given Cauchy data. General analytical solutions for this system are not known.
However, approximate solutions can be found through numerical simulations. We implemented this approach using a PDE solver~\cite{Mathematica}, using a finite element method. 
The simulations allowed us to study the Vortex and domain wall separately and their interactions. To approximate the initial conditions, we construct a Vortex-Coulomb Vacuum Layer configuration described below.

\begin{figure}
\centering
 \begin{subfigure}{0.4\textwidth} 
 \includegraphics[width=\textwidth]{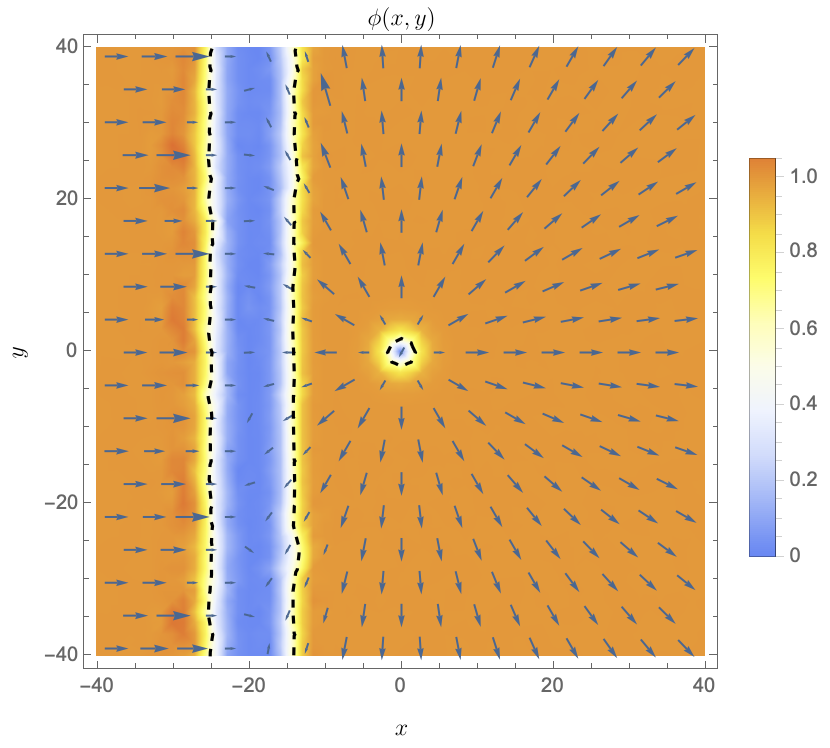}
 \end{subfigure}
\begin{subfigure}{0.4\textwidth} 		\includegraphics[width=\textwidth]{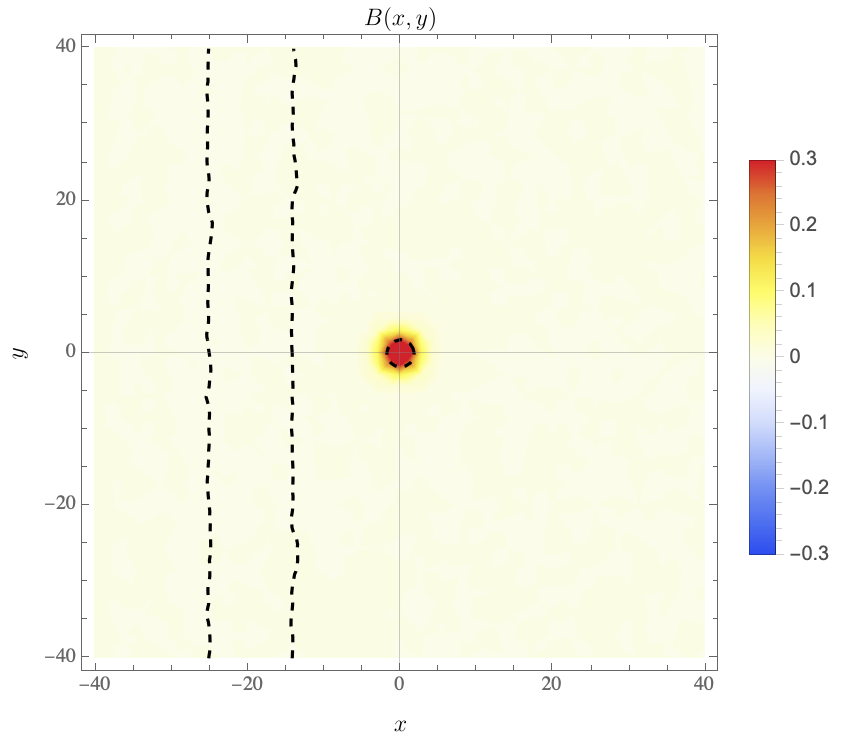}
	\end{subfigure}

	\caption{Vortex-Coulomb Vacuum Layer configuration $\phi_{\text{vo-dw}}(x,y)$ at $t=0$. Here $\nu=1$, $m_h=1$, $l=20$, $L_{\text{vd}}=20$, and $v=0.8$. On the top, the field $\phi$ is plotted. Observe the Coulomb vacuum layer centered at $x=-20$ and the vortex core at the origin. The phase $\text{Arg}(\phi)$  winds around the origin while for $x<-20$, the phase is approximately constant. Below, we plot the magnetic field, $B(x,y)$ corresponding to the gauge field configuration $A_{\text{vo-dw}}(x,y)$. The dashed black lines correspond to the points where $|\phi(x,y)|=v/\sqrt{3}$, where the maximum of the potential is attained.} 
	\label{fig:Vortex_CVL}
\end{figure}

Lets denote the vortex field configuration (\ref{eq:ansatz-phi}), and (\ref{eq:ansatz-A}) by $\phi_\text{vo}(x,y)$, and  ${A_i}_{\text{vo}}(x,y)$, respectively. On the other hand, the Coulomb vacuum layer configuration we will consider is given by $\phi_{(\nu,0,\nu)}(x)$--equation (\ref{eq:DW-Ansatz}).

For the Vortex-Coulomb Vacuum Layer configuration, we consider the following ansatz to approximate the initial conditions:
\begin{equation}
    \begin{aligned}
            \phi_{\text{vo-dw}}(x,y)=&\ \phi_{\left(\nu,0\right)}\left(\left(x+L_{\text{vd}}\right)+\frac{l}{2}\right)\\    &\ +\left(\frac{\phi_{(0,\nu)}\left(\left(x+L_{\text{vd}}\right)-\frac{l}{2}\right)}{\nu}\right)\phi_\text{vo}(x,y),
    \end{aligned}
    \label{eq:DW-Vo-Ansatz-phi}
\end{equation}
\begin{equation}
    {A_i}_{\text{vo-dw}}(x,y)=\left(\frac{\phi_{(0,\nu)}\left(\left(x+L_{\text{vd}}\right)-\frac{l}{2}\right)}{\nu}\right){A_i}_{\text{vo}}(x,y),
    \label{eq:DW-Vo-Ansatz-A}
\end{equation}
where $L_{\text{vd}}$ is the initial distance between the cores of the vortex and the Coulomb vacuum layer. For $L_{\text{vd}}\to\infty$, the fields configuration (\ref{eq:DW-Vo-Ansatz-phi}), and (\ref{eq:DW-Vo-Ansatz-A}) reproduce asymptotically the required initial conditions, as it is shown in the appendix \ref{sec:appendix2}.
More generally, we consider an initial relativistic velocity, $v$, of the Coulomb vacuum layer. For doing so, we boost the domain wall profiles as shown in equation (\ref{eq:DW-Vo-Boost}). We checked numerically that the field configuration (\ref{eq:DW-Vo-Boost}) satisfies approximately the Gauss constraint (\ref{eq:Gauss_Constraint}), allowing us to use it as an initial condition. An example of such initial conditions is shown in Figure \ref{fig:Vortex_CVL}

During the simulation of the time evolution, we observed the unwinding of the scalar vortex when it enters the Coulomb phase, producing two perturbations that travel along the $\phi_{(0,\nu)}$-domain wall. 
In addition, the magnetic field gets unconfined in the Coulomb phase producing radiation modes that are reflected by the $\phi_{(\nu,0)}$-domain wall.
Subsequently, most of this radiation is confined to the core of the layer. Below we present these results, first describing the time evolution for the field $\phi$ and afterward the evolution of $A_i$ and $B$. 
Additionally, the results of our numerical simulations can be visualized in the following 
\href{https://youtu.be/6VFgjXrUHq0}{video:} \url{https://youtu.be/6VFgjXrUHq0}.

\begin{figure*}
	\centering
    \includegraphics[width=0.8\textwidth]{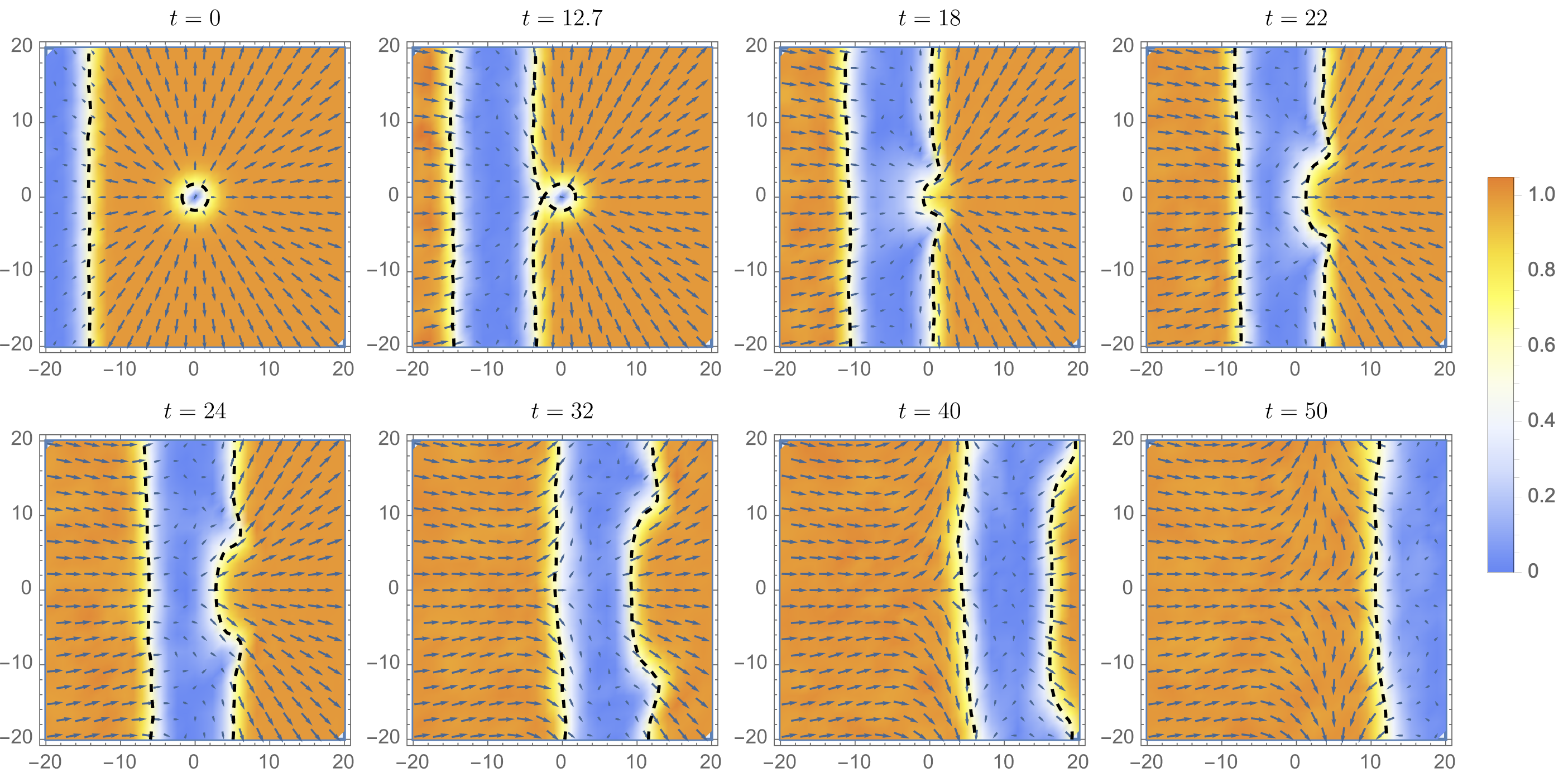}
	\caption{Time evolution of the field $\phi$ near the vortex position. At $t=0$, the initial vortex configuration is observed. On the other hand, at $t=50$, no winding of the phase is observed. The color represents $|\phi(x,y)|/v$. The blue regions correspond to the Coulomb phase, while the orange regions correspond to the Higgs phase. The dashed black lines correspond to the points where $|\phi(x,y)|=v/\sqrt{3}$. Observe at $t=12.7$, the moment the vortex enters the Coulomb vacuum layer, and two perturbations are generated on the wall. At subsequent times observe how these two perturbations travel along the wall in opposite directions.}
	\label{fig:phi_TE}
\end{figure*}

\textit{$\phi$ evolution:}  Figure \ref{fig:phi_TE} shows the time evolution of the scalar field, $\phi$. Recall that in the Higgs phase, the degree of freedom corresponding to $|\phi|$ becomes the  degree of freedom of the neutral scalar field $h$. On the other hand, in the Coulomb phase, $|\phi|$ is one of the two degrees of freedom of the complex field $\phi$. Observe the Wave modes corresponding to $h$ and $|\phi|$ that are generated as the domain wall interacts with the vortex. As the vortex encounters the Coulomb vacuum layer, two perturbations on the $(0,\nu)$-domain wall are produced, and they start propagating in opposite directions along the wall. 

If a domain wall evolution is tension-dominated, the evolution of perturbations on the wall is  effectively described by considering the thin-Wall approximation. Then the total energy will be proportional to the length (or area in $(3+1)$-dimensions) of the wall. The corresponding action is the Nambu-Goto action~\cite{Vilenkin:2000jqa}. 
\begin{equation}
    S=-\sigma\int d\mathcal{A},
\end{equation}
where $\sigma=\frac{m_h^2}{8\lambda}$ is the tension of the $(\nu,0)$-domain wall, and $d\mathcal{A}$ is the differential area of the world-sheet. The dynamics of such perturbations on domain walls have been extensively studied in~\cite{Blanco-Pillado:2022rad}. It can be shown that the speed of propagation of perturbations on the wall is $c=1$, as it is observed in our simulation.

\begin{figure*}
	\centering
    \includegraphics[width=0.8\textwidth]{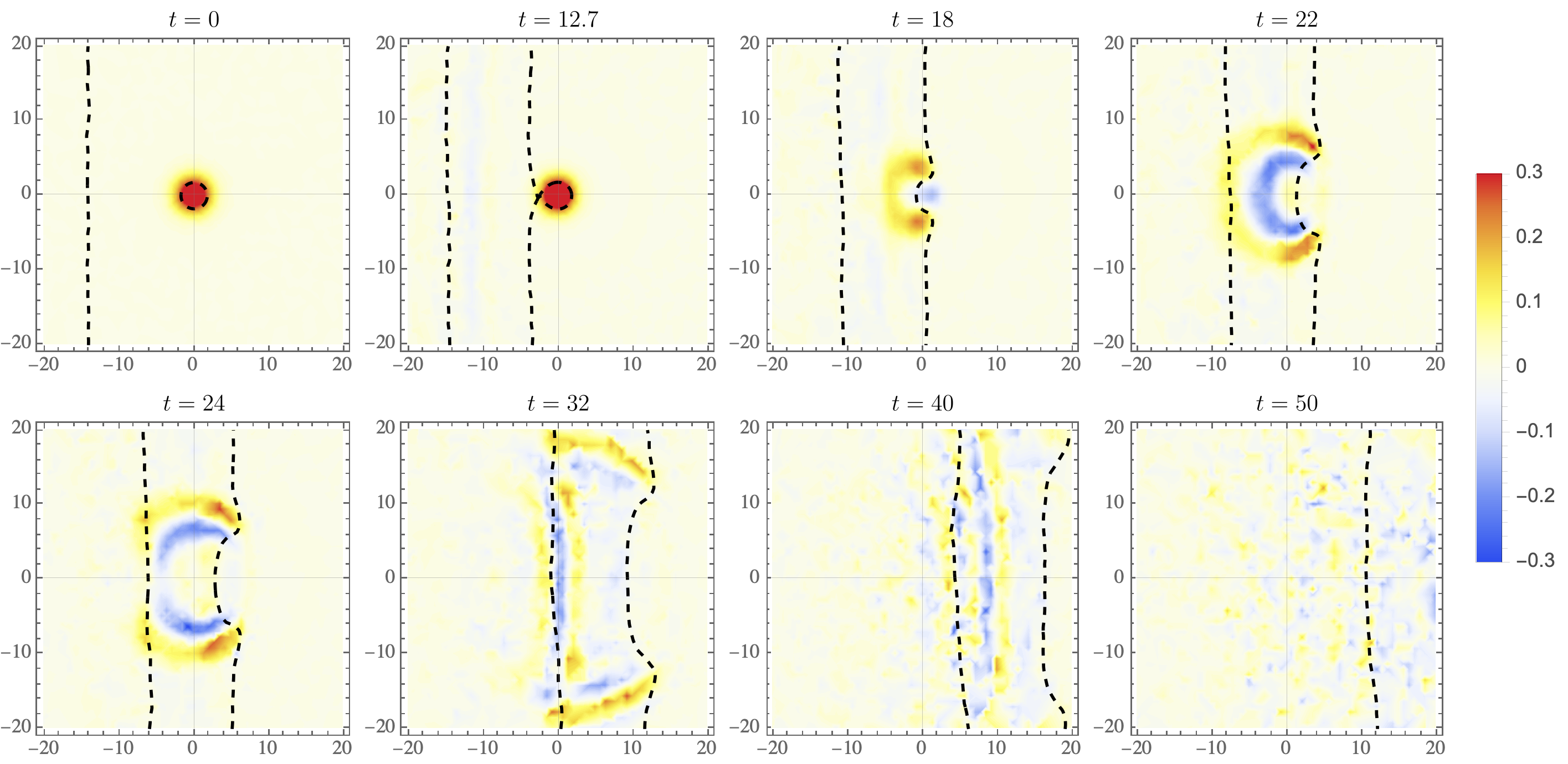}
	\caption{Time evolution of the magnetic field  $B$ near the vortex position. At $t=0$, the initial field configuration corresponds to the vortex configuration. As the Coulomb layer encounters the vortex, at $t=12.7$, the magnetic field starts to dissipate in the core of the layer. As a consequence, electromagnetic radiation is produced. Observe a hemispherical wavefront of such radiation at $t=22$. At $t=24$, the $(\nu,0)$-domain wall encounters the wavefront, and most of the electromagnetic radiation is reflected.} 
	\label{fig:B_TE}
\end{figure*}

\textit{$B$ evolution:} The time evolution of the field $A_i$ provides the time evolution of the magnetic field $B$, which is shown in Figure \ref{fig:B_TE}. As the vortex enters the Coulomb layer, the electric current $j^\mu$--that localizes the magnetic field on the core of the vortex--approaches zero. In fact, the electric current, $j^\mu=-i\left(\phi^*D^\mu \phi-(D^\mu \phi)^*\phi\right)$,
is proportional to the norm of the field $|\phi|$. Since $|\phi|\sim 0$ in the core of the Coulomb Vacuum Layer, then $|j^\mu|\sim 0$. Consequently, the field equations (\ref{eq:eom-A}) become approximately the Maxwell equations in (Coulomb) vacuum. Thus, the magnetic field dissipates in the core of the layer while producing electromagnetic radiation. This behavior is precisely observed at $t=40$ in Figure \ref{fig:B_TE}. Afterward, as the front of the electromagnetic radiation encounters the $(\nu,0)$-domain wall, it gets reflected. 

\section{Physics of erasure} 
In our numerical analysis, we observe that vortex never makes it to the other side of the layer. 
Once the vortex enters the layer, it unwinds and the magnetic flux spreads over,  suppressing the probability for the vortex to be recreated on the opposite side of the wall. In other words, we observe that the erasure happens with almost a unit probability. 
That is, it nearly saturates the unitarity of the evolution process. As we have done it previously for the erasure of the colliding monopole-anti-monopole pair~\cite{Dvali:2022vwh}, we give to the present phenomenon the interpretation in terms of a loss of coherence and an undersaturation by the vortex of the unitarity entropy bound in the sense of~\cite{Dvali:2020wqi}. 
 
 Namely, in the considered model, the vortex represents a coherent state of quanta of very low microstate degeneracy. 
Correspondingly,  its microstate entropy is much lower than  the maximal entropy permitted by the unitarity bound of 
~\cite{Dvali:2020wqi}. In the field configuration space, the vortex corresponds to a very special configuration in which the particles are arranged in a highly coherent way. 
It, therefore, covers a tiny fraction of all possible micro-states.  
In other words, the random field configurations of similar winding numbers and energy, which are supported by the layer, have much higher entropy.  
  
Of course, far away from the layer, the vortex corresponds to the lowest energy state for a given winding number. 
However, when the vortex collides with the layer, it produces waves. They have a double effect. First, they let the magnetic field spread out. 
Secondly, they carry away some coherence.
That is, the constituents of the vortex ``thermalize''. 
The  further time evolution is towards the more entropic state.
Entropically, it is much more convenient for the field to further spread out, rather than to compose a new vortex. 
Due to this, the recreation of the vortex is suppressed by the entropy factor.

Lifting the problem to one higher dimension, the same reasoning applies to the erasure of the cosmic string by a domain wall. 
Moreover, in a similar kinematic regime, the same outcome is expected in the case of a passage of the  QCD string through the deconfining domain wall in DS model~\cite{Dvali:1996xe}.  
The gluon electric flux, which outside of the wall is trapped in a tube,  upon colliding with the wall, spreads over it, as  described by Figure \ref{fig:StringsOnWall}. 
Again, the entropy argument tells us that the recreation of the string on the other side of the wall must be exponentially suppressed since the string carries a much lower entropy than the spread-out flux. 
As we already discussed,  due to quantum tunneling effects, inside the wall, the flux will not spread all the way to infinity but will become exponentially wide~\cite{Dvali:2007nm}. 
This does not make much difference for the erasure mechanism since the entropy of the spread-out flux is still much higher relative to a localized flux tube outside of the wall.  

Due to the analogy between the field-theoretic walls of~\cite{Dvali:1996xe} and $D$-branes~\cite{Dvali:2002fi}, the same outcome is expected in the interactions between $F$-strings and $D$-branes. 
 This can have implications for cosmology and the signatures of gravitational waves. 
 
 \section{Implication for brane-world scenarios} 

Here, we wish to discuss that the erasure of strings is a generic prediction for the brane-world scenarios, such as~\cite{Arkani-Hamed:1998jmv}. 
This is due to universal fundamental features of the phenomenon of the gauge field localization on the brane~\cite{Dvali:1996xe, Arkani-Hamed:1998jmv}.  
   
In brane-world scenarios, the massless gauge fields (such as the $U(1)$-photon) are localized on a $3$-brane (i.e., the brane with $3$ non-compact spatial world-volume dimensions, which we can label as $x_j, ~j=1,2,3$).  
This $3$-brane  is embedded in space with $n$ compact extra dimensions, with coordinates $y_a, ~ a=1,2,...,n$.   
    
This setup universally implies the existence of strings with $U(1)$ electric flux~\cite{Dvali:1996xe, Arkani-Hamed:1998jmv}. 
This follows from the gauge invariance of the $3+1$-dimensional $U(1)$-theory of a massless photon, which demands that the medium outside of the brane must confine the $U(1)$ electric flux into flux tubes.
In the DS model~\cite{Dvali:1996xe}, in which the localization mechanism was originally proposed, these are QCD flux tubes. 
However, the phenomenon is very general.  As explained there, the localization requires a dual Meissner effect, which implies confinement in the bulk. How this confinement is realized in each particular case, is not essential for the existence of strings. 
   
For example, in string theoretic realization~\cite{Antoniadis:1998ig} of brane-world model of~\cite{Arkani-Hamed:1998jmv}, the $3$-brane is a $D$-brane and the bulk flux tubes are the fundamental strings.
These can play the role of cosmic $F$- and $D$- strings~\cite{Copeland:2003bj, Dvali:2003zj}.

Thus, a generic brane-world setup, with localized photons, universally includes the strings in the form of flux tubes and the $3$-branes that deconfine this flux.  
Correspondingly, when a string meets the brane, it gets erased. 
     
In order to be more precise, let us consider a situation in which ``our'' $3$-brane  is placed at $y_a=0$ point in extra space. The brane supports a $U(1)$-gauge theory with the massless photon in the Coulomb phase. 
The electric flux of the same $U(1)$, in the bulk, is confined in flux tubes, which form strings. 
Now let us consider a bulk electric string that is extended along one of the world-volume dimensions, say $x_3$, and simultaneously is displaced from our brane in the extra space by a distance $r$. 
Let us assume that the string coordinates are $y_a=\delta_{a1}r$.  
For the world-volume observers of our $3$-brane, the string looks like an ordinary straight cosmic string extended along $x_3$-axes. This continues to be true up until the string touches our brane. At this point, the portion of the string, that enters the brane, gets erased. That is, the flux gets deconfined and  spreads out in the form of an electric field of a dipole, as it is depicted in Figure \ref{fig:StringsOnWall}. 
From the point of view of  a world-volume observer, the  string-wall junction points appear as electric charges. 
If the angle between the string and the 3-Brane is small, the charges will be pulled apart by the string tension. 
 
This break-up of the  cosmic string is very different from a more familiar break-up due to the nucleation of a pair of monopole and anti-monopole~\cite{Vilenkin:1982hm}.  
First, the monopole nucleation process is exponentially suppressed, whereas the erasure has a probability close to one.  
Secondly, in the case of monopoles, the endpoints do not create an electric field of a dipole. 
   
More importantly, in general, the bulk string is expected to enter our brane-world in multiple locations. 
So the break-up of the string will happen simultaneously in several places, leading to spectacular events accompanied by the emission of gravitational waves and electromagnetic radiation.  
They can be of observational importance and, in particular, are expected to give correlated signals. 

\section{Conclusion}
In this paper, the collision of a vortex with a layer of a Coulomb vacuum is simulated numerically, and it is found that none of the vortices cross the Coulomb vacuum layer.
We observe how the collision leads to the unwinding of the vortex and the un-confinement of the magnetic flux, which dissipates in the core of the layer. 
The same behavior of vortex unwinding was observed in different regimes of parameters of the model, different winding numbers $n$, and different widths of the Coulomb vacuum layer $l$.
Our results allow us to conclude that the mechanism of the erasure of defects
takes place in the model (\ref{eq:Lagrangian}).
The outcome is largely insensitive  to the parameters.

We explain the observed robustness of the erasure phenomenon by the  arguments that have been originally used in~\cite{Dvali:1997sa} in the context of $3+1$-dimensional wall/monopole system and were refined recently in~\cite{Dvali:2022vwh}  in terms of the low entropy of the vortex relative to the unitarity bound~\cite{Dvali:2020wqi}. 
  
These arguments suggest that the high probability of the erasure is due to the fact that upon the collision with the wall,  the vortex of the $2+1$-dimensional theory (or a string of the $3+1$ dimensional one) enter the deconfinement phase during which the flux spreads-out over the wall. 
This takes away some coherence required for the formation of a vortex (string). 
A further recreation of the vortex is exponentially suppressed due to its low microstate entropy. This diminishes the phase space for its production.  
Instead, the system evolves towards a much higher entropy state of the spread-out flux and waves.   
     
Since the argument is very general, it must equally apply both to cosmic strings which carry magnetic fluxes, as well as to QCD strings, which represent the electric flux tubes.
Thus, the QCD strings scattered at a layer of the deconfined phase, as is the case in the DS model 
~\cite{Dvali:1996xe}, is expected to be erased with almost unit probability. 
   
Judging from~\cite{Copeland:2003bj, Dvali:2003zj}, the same outcome is expected for the cosmic $F$-strings upon hitting the $D$-branes.  The further study of  such systems is interesting from fundamental as well as observational perspectives. In particular,  they can lead to distinct signatures of gravitational waves. 
Also,  preventing a cosmic string from passing through the brane can lead to a high concentration of energy in the wall vicinity,  which can result in a black hole formation in the spirit  of~\cite{Dvali:2021byy}.   
  
Finally, we have pointed out that due to the universal fundamental features of the DS gauge field localization mechanism on the brane~\cite{Dvali:1996xe, Arkani-Hamed:1998jmv}, the existence of erasing electric cosmic strings is a generic prediction of brane-world scenarios. 
Due to the $U(1)$ gauge invariance of the localized photon, such theories inevitably possess the bulk strings. 
Strings represent the electric flux of the ordinary photon, which gets confined into tubes away from  ``our'' brane.
This happens due to a dual Meissner effect in the bulk, which is a necessary condition for the  localization on the gauge field on the brane~\cite{Dvali:1996xe, Arkani-Hamed:1998jmv}.
 
Upon touching our brane-world, the cosmic strings get erased, breaking apart in multiple locations.  
This breakup is expected to result in spectacular events accompanied by gravitational and electromagnetic radiation of an observational interest.  
  
\section*{Acknowledgments} \label{sec:acknowledgements}

We thank Andrei Kovtun and Michael Zantedeschi for their helpful discussions. We also thank Alex Vilenkin for discussions on related topics.
This work was supported in part by the Humboldt Foundation under Humboldt Professorship Award, by the Deutsche Forschungsgemeinschaft (DFG, German Research Foundation) under Germany’s Excellence Strategy - EXC-2111 - 390814868, and Germany’s Excellence Strategy under Excellence Cluster Origins.

\bibliography{bibliography}
\appendix*
\section{Appendix}
\label{sec:appendix}
\subsection{Coulomb Vacuum Layer Decay}
\label{sec:appendix1}
\begin{figure}
	\centering
	\begin{subfigure}{0.4\textwidth} 
		\includegraphics[width=\textwidth]{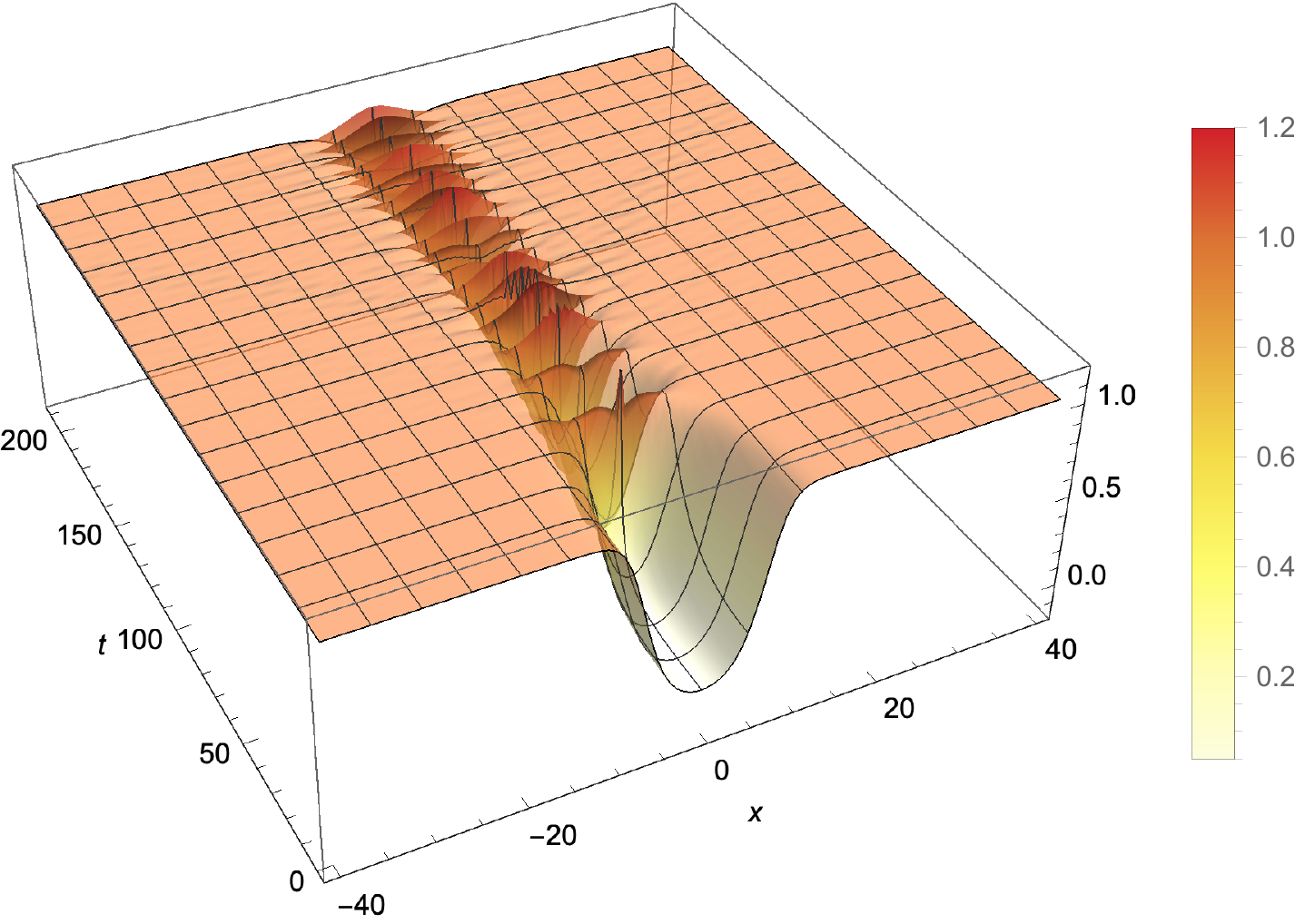}
		\caption{$\phi(t,x)$} 
	\end{subfigure}
	\begin{subfigure}{0.4\textwidth} 
 \includegraphics[width=\textwidth]{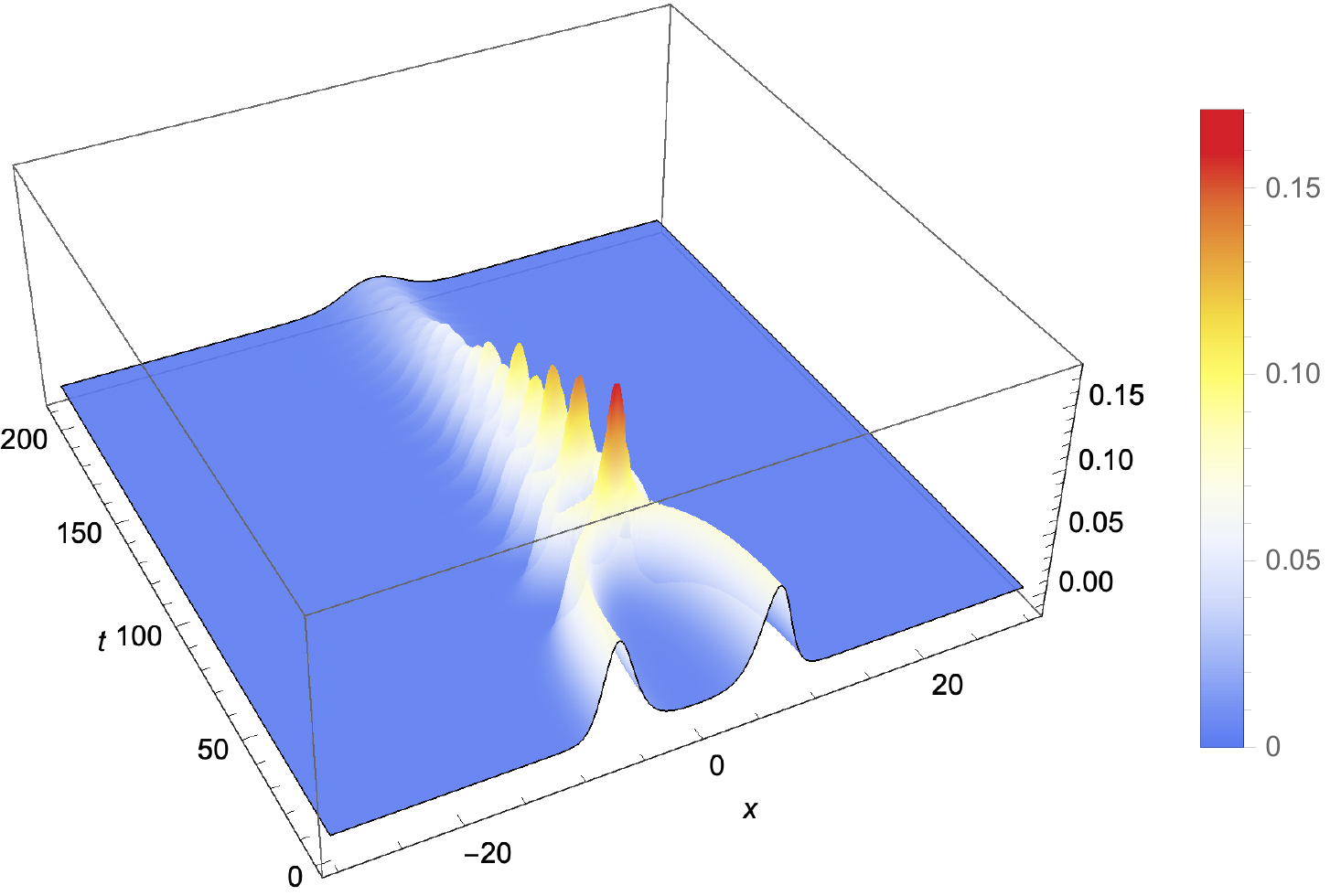}
		\caption{$\mathcal{E}[\phi(t,x)]$} 
	\end{subfigure}
	\caption{Time evolution of a $(\nu,0,\nu)$-domain wall. We observe the creation of a bion after the collision of the two Walls. (a) shows the time evolution of the field profile, while (b) shows the time evolution of the energy density.  At $t=0$, the core size of the Coulomb vacuum layer is $l=15 m_h^{-1}$.} 
	\label{fig:(N0N)-CVL-Decay}
\end{figure}

\begin{figure}
	\centering
	\begin{subfigure}{0.4\textwidth} 
 \includegraphics[width=\textwidth]{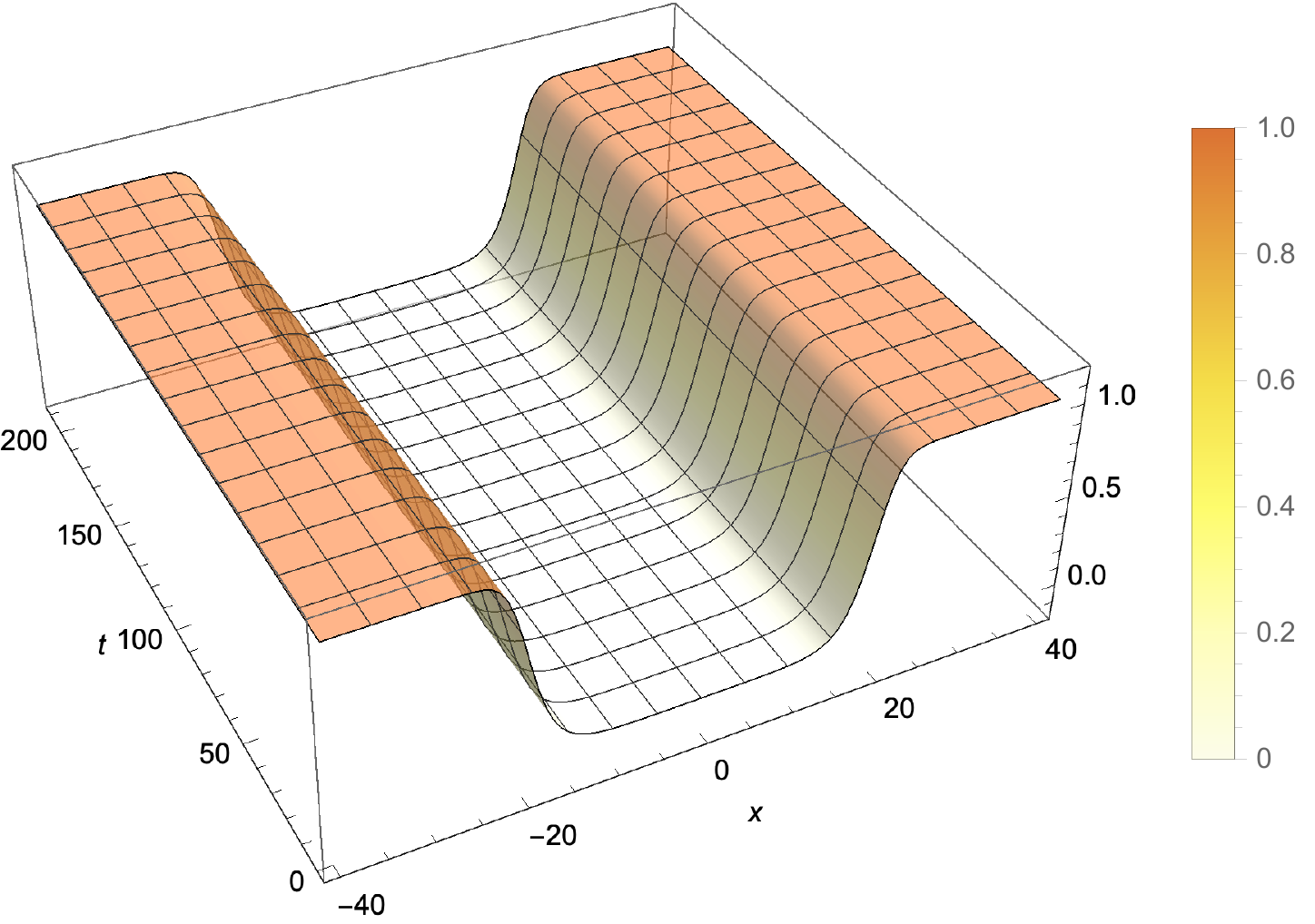}
		\caption{$\phi
		(t,x)$} 
	\end{subfigure}
	\begin{subfigure}{0.4\textwidth} \includegraphics[width=\textwidth]{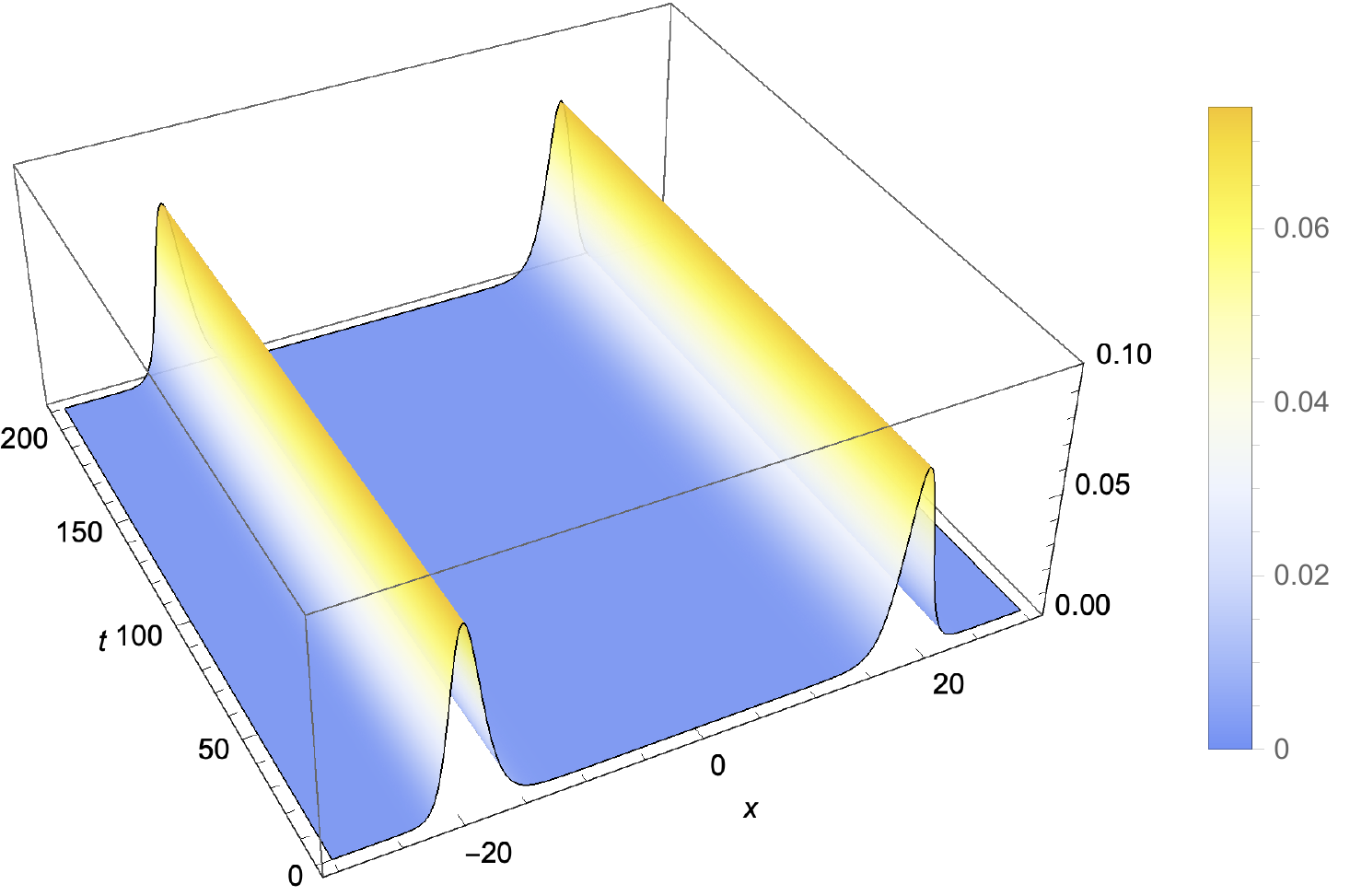}
		\caption{$\mathcal{E}[\phi
		(t,x)]$} 
	\end{subfigure}
	\caption{Time evolution of a of a long-lived $(\nu,0,\nu)$-domain wall. At $t= 0$, the core size of the  vacuum layer is $l= 40 m_h^{-1}$. The  time  evolution  of  the  field  profile  and  the  energy  density are shown in	(a) and (b), respectively.} 
	\label{fig:(N0N)-CVL-Long-Lived}.
\end{figure}

The Coulomb vacuum Layer configuration (\ref{eq:DW-Ansatz}) is not a solution to the static field equations, and consequently, it evolves in time and will decay.
This behavior can be observed in Figure \ref{fig:(N0N)-CVL-Decay}. We circumvent this by taking a larger initial distance $l$ so the vacuum layer can be considered long-lived for the time scales of our simulations. An example of such a case is shown in Figure \ref{fig:(N0N)-CVL-Long-Lived}. For this simulation, we have fixed the parameters $\nu=1$ and $\lambda=1/2$, corresponding to $m_h=1$.

\subsection{Initial Conditions}
\label{sec:appendix2}

For $L_{\text{vd}}\to\infty$, the fields configuration (\ref{eq:DW-Vo-Ansatz-phi}) and (\ref{eq:DW-Vo-Ansatz-A}) reproduce asymptotically the required Vortex-CVL initial conditions. To observe this limit, consider a region of space  $\mathcal{R}=[-L_{x},L_{x}]\times[-L_{y},L_{y}]$ such that 
$$ \frac{1}{m_h} < L_{y}\ll L_{x}\ll L_{\text{vd}}.$$ In the Limit $L_{\text{vd}}\to \infty$,  the field configuration for $x>-L_x$ approaches the vortex configuration

$$
\lim_{\substack{L_{\text{vd}}\to \infty \\ x>-L_x}}
 \phi_{\text{vo-dw}}(x,y) = \phi_\text{vo}(x,y),
$$
$$
\lim_{\substack{L_{\text{vd}}\to \infty \\ x>-L_x}}  {A_i}_{\text{vo-dw}}(x,y) = {A_i}_{\text{vo}}(x,y).
$$

On the other hand, if $x<-L_x$, the field configuration approaches a $(\nu,0,e^{i\alpha}\nu)$-domain wall configuration centered at $x=-L_{\text{vd}}$, i.e.
\begin{equation}
    \begin{split}
\lim_{\substack{L_{\text{vd}}\to \infty \\ x\ll-L_x}}  \phi_{\text{vo-dw}}(x,y) =& \phi_{\left(\nu,0\right)}\left(x'+\frac{l}{2}\right)\\ &+\phi_{(0,\nu)}\left(x'-\frac{l}{2}\right) e^{i\alpha(x,y)},    
\end{split}
\end{equation}
\begin{equation}
    \lim_{\substack{L_{\text{vd}}\to \infty \\ x\ll-L_x}}  {A_i}_{\text{vo-dw}}(x,y) < {A_i}_{\text{vo}}(-L_x,y)\sim 0,
\end{equation}
where $x'=x+L_{\text{vd}}$, and $e^{i\alpha(x,y)}=\left(\frac{x+i y}{\sqrt{x^2+y^2}}\right)^n$.

Moreover, if $|y| \ll |x|$, then $e^{i\alpha(x,y)}\sim (-1)^n$. Thus,
\begin{equation}
    \begin{split}
        \lim_{\substack{L_{\text{vd}}\to \infty \\ x<-L_x \\|y|< L_y}}  \phi_{\text{vo-dw}}(x,y) 
        =& \phi_{\left(\nu,0\right)}\left(x'+\frac{l}{2}\right) \\
        &+\phi_{(0,(-1)^n\nu)}\left(x'-\frac{l}{2}\right)\\
        =&\phi_{\left(\nu,0,(-1)^n\nu\right)}(x').
    \end{split}
\end{equation}

We conclude that the field configuration $(\phi_{\text{vo-dw}}(x,y),{A_i}_{\text{vo-dw}}(x,y))$, in the limit $L_{\text{vd}}\to \infty$, approaches a Vortex configuration near the origin, and  a Coulomb vacuum layer configuration for $x\sim-L_{\text{vd}}$. 
In addition, motivated by the fact that domain walls are generally very highly energetic objects moving through space, we consider an initial relativistic velocity, $v$, of the Coulomb vacuum layer. To do it, we boost the domain walls profiles as 
\begin{equation}
    \begin{aligned}
     \phi
     (t,x,y)=
        &\ \phi_{\left(\nu,0\right)}\left(\gamma\left(x + L_{\text{vd}}-v t +\frac{l}{2}\right)\right)\\
        &\ +\left(\frac{\phi_{(0,\nu)}\left(\gamma\left(x+L_{\text{vd}}-v t -\frac{l}{2}\right)\right)}{\nu}\right)\phi_\text{vo}(x,y),\\
    {A_i}
    (t,x,y)=
        &\ \left(\frac{ \phi_{(0,\nu)}\left(\gamma\left(x+L_{\text{vd}}-v t -\frac{l}{2}\right)\right)}{\nu}\right){A_i}_{\text{vo}}(x,y),
    \end{aligned}
    \label{eq:DW-Vo-Boost}
\end{equation}
where $\gamma=\frac{1}{\sqrt{1-v^2}}$ is the Lorentz factor. Figure \ref{fig:Vortex_CVL} shows an example of the field configuration $\phi_{\text{vo-dw}}(t,x,y)$, and ${A_i}_{\text{vo-dw}}(t,x,y)$ at $t=0$, respectively.

\subsection{Unwinding of the Vortex: Time evolution of \texorpdfstring{$n$}{n}}

\begin{figure}
	\centering
 	\includegraphics[width=0.4\textwidth]{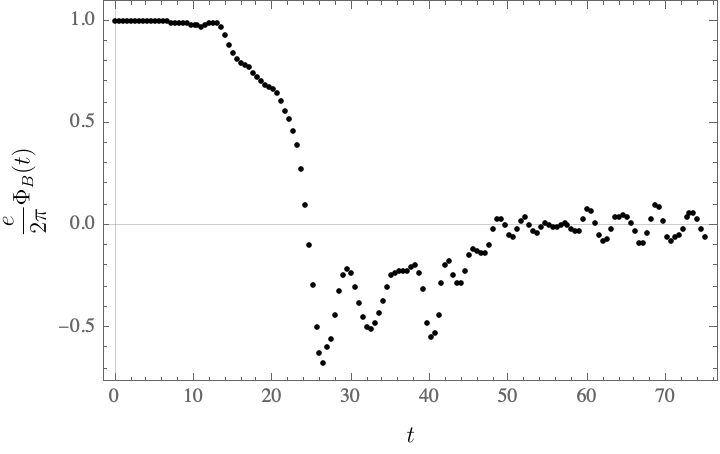}
	\caption{Time evolution of the magnetic Flux $\Phi_B$.} 
	\label{fig:Magnetic-Flux-TE}
\end{figure}

To describe the unwinding process of the vortex, we have numerically computed the time evolution of the winding number $n$ near the vortex position by two different methods. As we described before, the winding number is proportional to the magnetic flux $\Phi_B$--see equation (\ref{eq:winding_B}). We used this first method to compute the winding number over a finite region $C_r$. 
As shown in Figure \ref{fig:Magnetic-Flux-TE}, for $r=10$, the magnetic flux
is initially constant for the Vortex configuration we are considering. As the vortex is swept away by the Coulomb vacuum Layer, The flux stops being localized in the region and thus decreases and eventually tends to $0$. Figure \ref{fig:Magnetic-Flux-TE} shows precisely this behavior for the time evolution of $n(t)$. From the simulation results described above, this is to be expected since the magnetic field dissipates in the layer, and its flux is not localized around the origin any more.

\begin{figure}
\centering 
\includegraphics[width=0.4\textwidth]{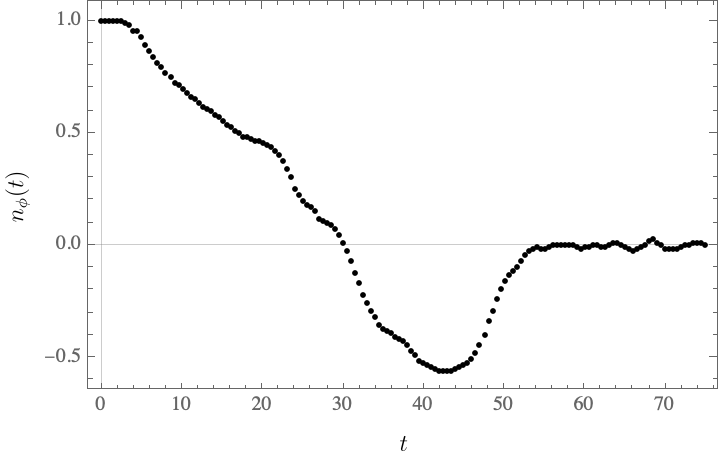}
	\caption{Time evolution of winding number $n_\phi$ computed in a finite region around the origin.} 
	\label{fig:Winding-Number_phi_TE}
\end{figure}

\begin{figure*}
\centering
    \includegraphics[width=0.9\textwidth]{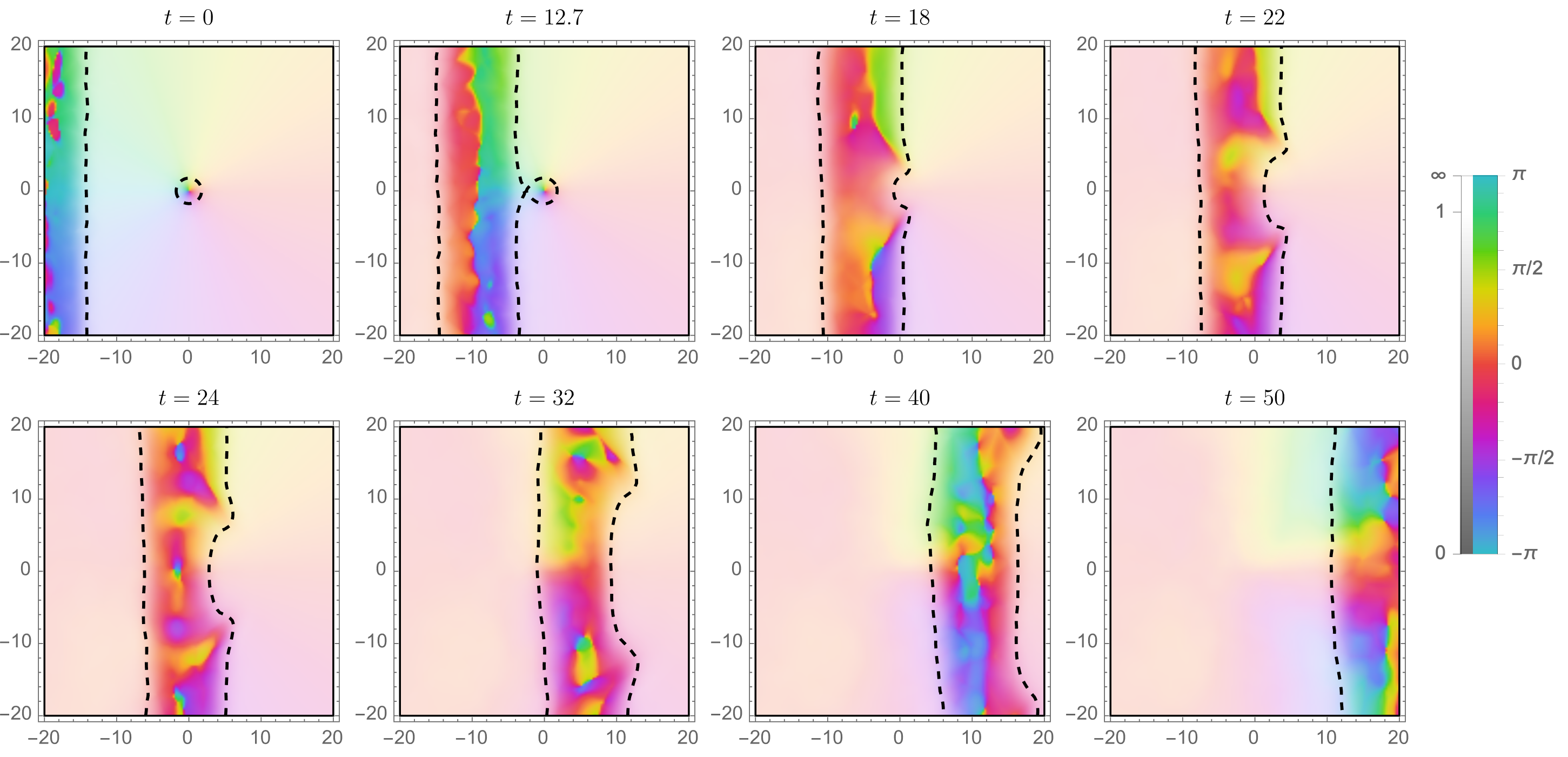}
	\caption{Time evolution of the winding of field $\phi$ for the same simulation shown in Figures \ref{fig:phi_TE} and \ref{fig:B_TE}. The color represents the phase $\text{Arg}(\phi)$, while the darkness measures the norm $|\phi|$. At $t=0$, note the winding $n=1$ around the vortex. Once the layer swaps the vortex, the phase $\text{Arg}(\phi)$ unwinds, and the vortex is erased.} 
	\label{fig:Unwinding_TE}
\end{figure*}

\begin{figure*}
\centering
    \includegraphics[width=0.9\textwidth]{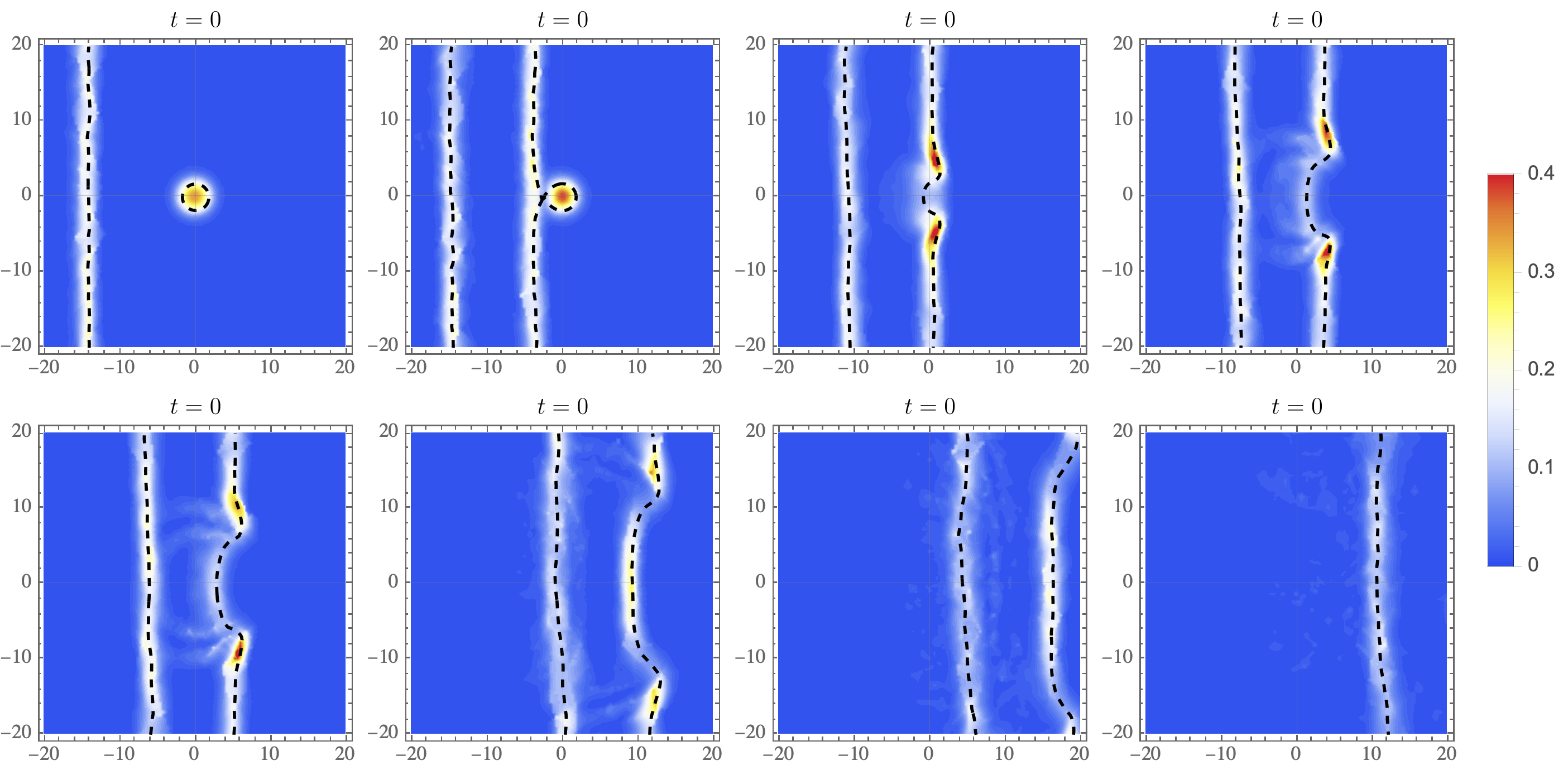}
	\caption{Time evolution of the total energy density for the simulation shown in Figures \ref{fig:phi_TE} and \ref{fig:B_TE}. Observe that most of the vortex energy is carried away by the modes propagating along the wall} 
	\label{fig:Energy_TE}
\end{figure*}

The second method we used to compute the winding number is in terms of the scalar field. From equation (\ref{eq:winding_Phi}), we define $n_{\phi}$ as
\begin{equation}
    n_\phi= \frac{1}{2\pi i \nu^2} \oint_{C_r} d x^i \frac{1}{2}\left( \phi^*\partial_i\phi-\phi\partial_i\phi^*\right),
\end{equation} 
where  $r$ is a finite radius. Figure \ref{fig:Winding-Number_phi_TE} shows the time evolution of $n_\phi(t)$ for $r=10$. We observe that $n_{\phi}=1$ at $t=0$, as it corresponds to the vortex configuration. As the vortex approaches the Coulomb vacuum layer,  $n_\phi$ decreases and becomes negative. To understand this behavior, let us consider the case $t\sim18$. Notice that $n_\phi(18)\sim0.5$, and that the $(0,\nu)$-domain wall is localised approximately at $x=0$--see Figure \ref{fig:phi_TE}. 
As a first approximation, in the boundary of the region of integration $C_r$, 
\begin{equation}
\phi(t=18,x,y)\sim \Theta(-x)\nu e^{i n \theta}.  
\end{equation}
Thus the integral $n_{\phi}(t=18)\sim n/2=0.5$. As the Coulomb vacuum layer passes over the origin, the winding number $n_\phi$ is not well defined until the $(\nu,0)$-domain wall passes over the origin. As we mentioned before, the phase near the origin becomes $\text{Arg}(\phi)\sim 0$; thus, as a first approximation, in the region of integration $C_r$: $\phi(t,x,y)\sim \nu$, for $t\gtrsim50$. Thus the winding number $n_\phi\sim0$. From the previous results, We conclude that locally the vortex is unwinded once it is swept by the Coulomb Vacuum Layer.  The previous discussion and time evolution can be observed in more detail in Figures \ref{fig:Unwinding_TE} and  \ref{fig:Energy_TE}, where we plotted the $\text{Arg}(\phi)$ and the energy density during the unwinding process, respectively. Additionally, the results of our numerical simulations can be visualized in the following 
\href{https://youtu.be/6VFgjXrUHq0}{video:} \url{https://youtu.be/6VFgjXrUHq0}.

\end{document}